# UNIVERSITY OF READING

## Department of Meteorology

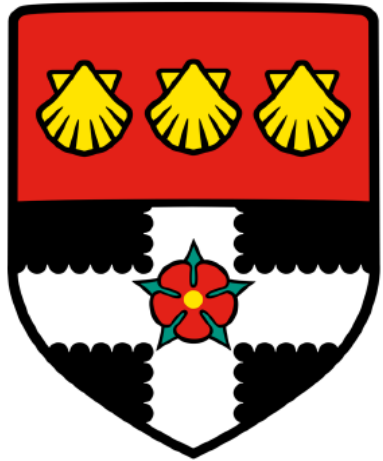

## Using Generative Models to Produce Realistic Populations of the United Kingdom Windstorms

Etron, Yee Chun TSOI

A dissertation submitted in partial fulfilment for the degree of
Master of Science in Applied Meteorology

August 2024

# ABSTRACT

Windstorms significantly impact the UK, causing extensive damage to property, disrupting society, and potentially resulting in loss of life. Accurate modelling and understanding of such events are essential for effective risk assessment and mitigation. However, the rarity of extreme windstorms results in limited observational data, which poses significant challenges for comprehensive analysis and insurance modelling. This dissertation explores the application of generative models to produce realistic synthetic wind field data, aiming to enhance the robustness of current CAT models used in the insurance industry.

The study utilises hourly reanalysis data from the ERA5 dataset, which covers the period from 1940 to 2022. Three models, including standard GANs, WGAN-GP, and U-net diffusion models, were employed to generate high-quality wind maps of the UK. These models are then evaluated using multiple metrics, including SSIM, KL divergence, and EMD, with some assessments performed in a reduced dimensionality space using PCA.

The results reveal that while all models are effective in capturing the general spatial characteristics, each model exhibits distinct strengths and weaknesses. The standard GAN introduced more noise compared to the other models. The WGAN-GP model demonstrated superior performance, particularly in replicating statistical distributions. The U-net diffusion model produced the most visually coherent outputs but struggled slightly in replicating peak intensities and their statistical variability.

This research underscores the potential of generative models in supplementing limited reanalysis datasets with synthetic data, providing valuable tools for risk assessment and catastrophe modelling. However, it is important to select appropriate evaluation metrics that assess different aspects of the generated outputs. Future work could refine these models and incorporate more complex architectures to further enhance their accuracy in wind field simulations and explore their applicability to other meteorological variables and modelling contexts.

# CONTENTS

## ACRONYMS AND ABBREVIATIONS

**CAT:** catastrophe

**CPU:** Central processing unit

**CUDA:** Compute Unified Device Architecture

**DCNN:** Deep convolutional neural network

**ECMWF:** European Centre for Medium-Range Weather Forecasts

**EMD:** Earth mover's distance

**ETC:** Extratropical cyclone

**FID:** Fréchet inception score

**GAN:** Generative adversarial network

**GPU:** Graphics processing unit

**IID:** Independent and identically distributed

**KL:** Kullback-Leibler

**LeakyReLU:** Leaky rectified linear unit

**McGAN:** Mean and covariance feature matching generative adversarial network

**MI:** Mutual information

**ML:** Machine learning

**NLPD:** Normalised Laplacian pyramid distance

**PC:** Principal component

**PCA:** Principal component analysis

**RAM:** Random-access memory

**SSI:** Storm severity index

**SSIM:** Structural similarity index measure

**UK:** United Kingdom

**VAE-GAN:** Variational autoencoder – generative adversarial network

**WGAN-GP:** Wasserstein generative adversarial network – gradient penalty

# 1. Introduction

*a.  Motivation*

Windstorms are among the most significant natural hazards in the United Kingdom (UK), with far-reaching impacts on society, infrastructure, and the economy (Baker & Lee, 2008). These severe weather events can cause extensive property damage, disrupt transportation networks, and lead to substantial financial losses. Historical records highlight that major windstorms such as the Great October Storm of 1987, the Burns Day storm in 1990, and the St. Jude storm in 2013, could collectively result in billions of pounds in insurance claims and considerable disruptions to daily life (Brooks, 2017; Dlugolecki, 1992; Hillier, 2017). These financial impacts underscore the critical need for an accurate understanding of windstorm frequency and severity.

Understanding the return period of extreme windstorms is essential for accurate risk assessment as it helps estimate the likelihood of future events and aids in preparation. However, one of the primary challenges in addressing the impact of windstorms is the inherent rarity of these extreme events. While moderate windstorms occur relatively frequently in the UK, extreme windstorms that have the potential to cause catastrophic damage are rare. This scarcity of data and the finite observational records pose significant challenges for meteorologists and risk assessors. Traditional reanalysis datasets are limited in providing the full spectrum of windstorm behaviour, particularly for extreme events (Priestley et al., 2018). Consequently, existing risk models often lack the robustness to accurately assess the risk associated with these infrequent but high-impact events (Sobel & Tippett, 2018; Yang et al., 2023).

The insurance industry relies on accurate risk assessments to set premiums and ensure financial stability in natural disasters (Collier et al., 2009). Catastrophe models are crucial in estimating the likelihood and severity of various scenarios based on historical data (Clark, 2002). However, the limited availability of data on extreme windstorms and the computational expensiveness to build very large hazard datasets with regional climate models hinders the development of robust CAT models. For instance, existing models may not adequately capture the tail-end risks associated with extreme events, leading to insufficient preparedness in financial planning (Sobel & Tippett, 2018).

Recently, advancements in synthetic data generation have offered feasible solutions to these limitations (Marwala et al., 2023). Generative models such as Generative Adversarial

Networks (GANs) and diffusion models have shown potential in producing realistic synthetic data. By generating high-quality, diverse wind field data that accurately reflects the characteristics of real windstorms, this study seeks to supplement existing reanalysis datasets, providing a more reliable basis for risk assessment and catastrophe modelling. Integrating synthetic data generated by generative models can enhance the robustness of catastrophe models, allowing for more accurate predictions of extreme windstorm events and their potential impacts (Hallegatte, 2007).

The potential benefits of this research extend beyond the insurance industry. Improved windstorm knowledge can aid emergency management agencies in developing more effective mitigation and response strategies. By better understanding the potential severity and frequency of extreme windstorms, authorities can enhance infrastructure resilience, improve disaster preparedness, and ultimately reduce the societal and economic impacts of these hazardous events. Furthermore, this approach holds promise for generating synthetic datasets for other data-driven models and weather prediction systems, opening new avenues for research and application in meteorology (Mukkavilli et al., 2023).

*b.* *Windstorms in the United Kingdom (UK)*

Windstorms, also known as extratropical cyclones (ETCs) or cyclonic systems, are a frequent and significant natural hazard in the UK. The geographical location of the country exposes it to frequent Atlantic ETCs, particularly during the winter. These storms bring strong winds, heavy rainfall, and occasionally snow, leading to various forms of societal and economic impact.

1) Historical Impacts

Historical records underscore the destructive power of several significant windstorms that have struck the UK. One of the most notable examples is the Great Storm of 1987, characterised by hurricane-force winds that caused extensive damage across southern England. This storm resulted in the loss of 18 lives, uprooted approximately 15 million trees, and caused an estimated 1.5 billion pounds in damage based on the currency value at the time (Jones, 2017; Owen, 2021). When adjusted for inflation, recent estimates suggest the cost of this storm would be approximately 9 billion pounds today (Cusack, 2023).

Another catastrophic event was the Burns Day Storm in January 1990, which brought winds exceeding 30 m s$^{-1}$. The storm claimed 47 lives and caused 3.37 billion pounds in damages at the time (McCallum, 1990). More recently, the St. Jude storm in October 2013 demonstrated the ongoing threat posed by these cyclonic systems. With wind gusts reaching up to 44 m s$^{-1}$, this storm resulted in the loss of five lives and significant disruption to transportation and power networks due to falling trees, with insurance losses estimated at 500 million pounds at the time (Brooks, 2017; Owen, 2021).

Following the St. Jude Storm, the winter of 2013-2014 brought a series of severe storms that compounded the financial burden on the UK insurance industry (Brown, 2014). These storms, combined with flooding, resulted in extensive damage to infrastructure, housing, and commercial properties. The financial cost of these storms was significant, with insurers having paid out more than 1.1 billion pounds in claims at the time (Kendon & McCarthy, 2015).

In addition to the direct financial costs, windstorms can cause profound societal disruption. Transportation networks, including roads, railways, and airports, are often affected, leading to delays and cancellations that impact daily life and economic activities. Power outages caused by damaged infrastructure can leave thousands of homes and businesses without electricity for extended periods, disrupting daily life and delaying emergency response (Dlugolecki, 1992). The widespread damage to infrastructure and natural landscapes often takes years to fully recover from, further adding to the long-term economic burden. These costs highlight the critical need for resilient infrastructure and effective disaster preparedness strategies.

2)    REGIONAL VARIATIONS AND SEASONAL TRENDS

The impact of windstorms varies across different regions of the UK. Coastal areas and regions with higher elevations, such as Scotland and Wales, tend to experience stronger wind than inland areas due to their lower surface roughness. Apart from topography, the variability in the wind speed during windstorms is influenced by several factors, including geographical location and the jet stream (Feser et al., 2015).

Most windstorms affecting the UK originate from the west and are driven by prevailing westerly winds and the jet stream, a fast-flowing region of the upper troposphere that can steer storm systems (Harding et al., 2009). During the winter months, from October to March, the jet stream is typically strongest and positioned over the North Atlantic, creating favourable conditions for the development and passage of cyclonic systems towards the UK (Mitchell-Wallace et al., 2017). These winter storms are often more intense due to the steeper temperature gradients between the polar and mid-latitude air masses, which fuel the development of strong storm systems (Feser et al., 2015).

Certain areas in the UK are more prone to severe impacts due to local geographical and climatic conditions. For instance, the southwest of England and Wales often face significant wind impacts due to their exposure to incoming Atlantic storms. Scotland, particularly the western and northern parts, experiences high wind speeds and frequent storms because of its topography and location relative to the typical storm tracks (Lamb & Frydendahl, 1991). These regions have shorter return periods for high-intensity storms, necessitating more robust emergency preparedness plans.

3)     Factors Influencing Windstorm Impact

Several factors contribute to the impact of windstorms in the UK. These include the intensity and path of storms, secondary weather phenomena such as heavy rainfall and storm surges, and the resilience of infrastructure and buildings. Some studies have also highlighted the role of climate change in potentially altering the frequency and intensity of windstorms, with some projections suggesting an increase in severe wind events in the future due to global warming (Dale et al., 2001; Peterson, 2000).

The most damaging elements of these storms include the peak wind gusts, which can cause structural damage, and the associated heavy rainfall that can result in significant flooding (Gliksman et al., 2023). Storm surges driven by strong winds pushing seawater onto the land can also contribute to coastal flooding and erosion (Burkett & Davidson, 2012). The combination of these factors creates a complex risk landscape and robust risk modelling approaches are necessary to accurately assess and mitigate the impacts of UK windstorms.

c.     *Importance of Catastrophe (CAT) Models*

1)     Role in Risk Assessment

Catastrophe (CAT) models are indispensable tools in the insurance and reinsurance industries. They provide a structured approach to estimating potential losses from extreme events such as cyclones, earthquakes, and floods. By integrating diverse data sources and applying sophisticated algorithms, CAT models play a pivotal role in understanding and managing the risks associated with natural catastrophes (Mitchell et al., 2017).

CAT models are central to risk assessment, which involves quantifying the likelihood and potential impact of natural disasters. This approach enables insurers and reinsurers to determine premiums, manage risk portfolios, support underwriting decisions, and facilitate reinsurance (Cummins & Trainar, 2009). By understanding the risk profile of different regions and assets, insurers can set premiums that accurately reflect the potential for loss, ensuring financial stability (Cummins et al., 1993).

2)  COMPONENTS OF CATASTROPHE (CAT) MODELS

A typical CAT model consists of several interconnected components (Fig. 1), each contributing to the overall risk assessment (Khanduri & Morrow, 2003; Mahdyiar & Porter, 2005; Zanardo & Salinas, 2022). The hazard module simulates the physical characteristics of the catastrophe event, such as the intensity and geographical distribution of a windstorm. The exposure module represents the assets at risk, including buildings, infrastructure, and other properties. The vulnerability module estimates the potential damage to these assets based on their exposure to the hazard. Finally, the financial module calculates the financial impact, including insured and uninsured losses, and applies various insurance and reinsurance policies to distribute the losses appropriately. These components work together to comprehensively assess potential losses, helping insurers understand their exposure and make informed decisions about risk management.

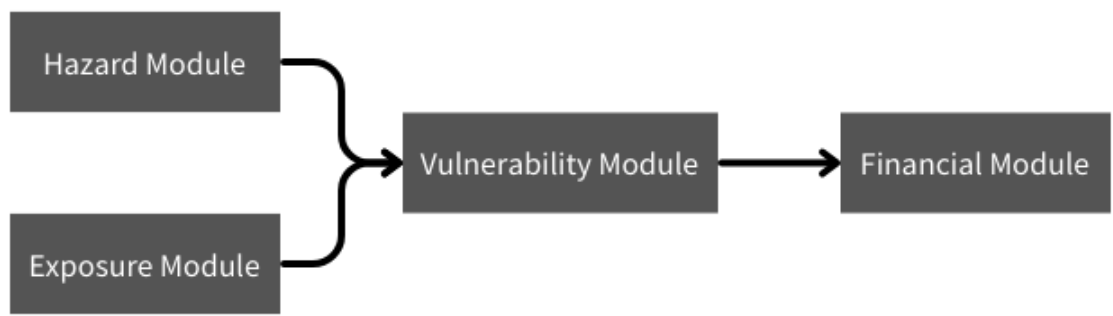

Figure 1. Components of a catastrophe risk model.

3)  CHALLENGES AND LIMITATIONS

The CAT model plays a crucial role in assessing windstorm risks but faces several challenges and limitations. A significant challenge is the uncertainties in modelling damages to buildings in response to given wind speeds in the vulnerability module, mainly due to the complex relationship between wind speeds and structural damages (Yum et al., 2021). Enhancing the resolution of the hazard module could allow more accurate characterisation of wind speeds at specific locations, which could improve the accuracy of vulnerability assessments.

Furthermore, there are uncertainties involved in modelling extremes, particularly in inferring the tail-end of the distribution from a limited amount of historical data (Mahdyiar & Porter, 2005). Extreme events are rare, and historical data recorded from the early 20[th] century may be insufficient when modelling these extremes in a spatially correlated manner, which is necessary for large-scale risk assessments. Additionally, changing climate patterns introduce additional uncertainty, making it difficult to rely solely on historical data for future risk assessments (Lloyd & Shepherd, 2020).

4)     INTEGRATION OF SYNTHETIC DATA IN CATASTROPHE (CAT) MODELS

The integration of statistical approaches offers promising solutions to some existing challenges by generating synthetic data that can supplement historical records (Sampson et al., 2014; Golnaraghi et al., 2018). This research explores and validates the potential of these advanced techniques. By leveraging generative models, the study aims to create a comprehensive dataset that captures a broader spectrum of windstorm scenarios, including extreme events that are underrepresented in reanalysis data.

d.     *Machine Learning in Meteorology*

The application of machine learning (ML) in meteorology has revolutionized various meteorological domains, including weather forecasting, climate modelling, and environmental monitoring (Bochenek & Ustrnul, 2022; Hino et al., 2018; Sun et al., 2021). By leveraging vast amounts of data, ML models can capture intricate patterns and relationships that traditional methods may miss (Brajard et al., 2021; Cho et al., 2020; Han et al., 2021). For instance, GraphCast from Google DeepMind has shown enhanced accuracy in global weather forecasting with a purely data-driven approach using advanced graph neural network (GNN) architectures (Lam et al., 2022).

In nowcasting, which involves short-term weather predictions, ML techniques combining with observations like radar and satellite images have improved storm development predictions in near real-time, providing timely and accurate forecasts for very short-term weather conditions (Woo & Wong, 2017; Shi et al., 2015). Additionally, ML models have been applied to predict different weather phenomena and climate model downscaling, contributing to better disaster preparedness and mitigation (Chapman et al., 2019; Rampal et al., 2024; Tsoi et al., 2023).

While ML has revolutionized meteorology and many other fields, challenges remain, particularly regarding data quality and model interpretability. Accurate ML predictions require large, high-quality datasets, and biases in the data can lead to errors. Additionally, the black-box nature of ML models can hinder their interpretability, making it difficult for meteorologists and other stakeholders to trust the predictions fully (Dueben et al., 2022).

Nevertheless, the future of ML in meteorology looks promising, with ongoing research aimed at integrating ML with physical models to leverage the strengths of both approaches. Advancements in computational power and explainable ML techniques will make ML models more transparent and trustworthy, facilitating their adoption in operational meteorology (Chen et al., 2023).

*e.    Generative Models: An Overview and Challenges*

Generative models are a powerful class of unsupervised machine learning algorithms designed to generate new data samples that resemble a given dataset (James et al., 2023). Unlike traditional discriminative models, which aim to classify or predict outcomes, unsupervised algorithms learn the underlying distributions within the data, enabling them to create realistic and diverse samples (Goodfellow et al., 2020).

1)    GENERATIVE ADVERSARIAL NETWORKS (GANs)

A neural network consists of layers of interconnected nodes known as neurons, where each connection has an associated weight (Aggarwal, 2018). Neural networks are designed to recognise patterns and learn from data through a training process, where the network weights are adjusted to minimise losses in its predictions. Generative Adversarial Networks (GANs), initially introduced by Goodfellow et al. (2014), consist of two neural networks, a generator and a discriminator, that are trained simultaneously through a process known as adversarial training.

The generator network $G$, creates synthetic data samples $G(z)$ from random noise $z$ in a latent space. The noise is typically drawn from a prior distribution $p_z(z)$, such as a uniform or normal distribution, which provides a source of randomness for generating diverse samples. The goal of the generator is to produce samples that are indistinguishable from actual data to the discriminator. On the other hand, the discriminator $D$ evaluates these synthetic samples $G(z)$ against actual data samples, denoted as $x$. It takes the samples as inputs and tries to determine whether each sample is real from the actual data distribution $p_{data}(x)$. The goal of the discriminator is to correctly classify the inputs as real ($D(x)$, labelled as 1) or fake ($D(G(z))$, labelled as 0).

During training, the generator tries to create better and more realistic data to fool the discriminator, while the discriminator continuously improves its ability to distinguish between real and synthetic data. This adversarial process continues until the generated samples are so realistic that the discriminator can no longer reliably tell the difference between real and generated data. This creates a two-player minimax game described by the following objective function (Goodfellow et al., 2014; Creswell et al., 2018):

$$min_G max_D V(D, G) = \mathbb{E}_{x \sim p_{data}(x)}[\log D(x)] + \mathbb{E}_{x \sim p_z(z)}\left[\log\left(1 - D(G(z))\right)\right]. \quad (1)$$

In this equation, $V(D, G)$ represents the value function that the GANs aim to optimise in this minimax game. The term $\mathbb{E}_{x \sim p_{data}(x)}[\log D(x)]$ represents the expected value of the logarithm of the discriminator output for real data, which aims to improve the skills of the discriminator in identifying real data. The term $\mathbb{E}_{x \sim p_z(z)}\left[\log\left(1 - D(G(z))\right)\right]$ represents the expected value of the logarithm of one minus the discriminator output for generated data, which aims to limit the ability of the discriminator to correctly identify the generated data as fake.

## 2)    DIFFUSION MODELS

Diffusion models are another type of generative model that has gained attention for their ability to generate high-quality data samples. These models are inspired by the diffusion process in physics, which generates new data samples by learning to reverse a gradual noise addition (Croitoru et al., 2023). A diffusion model typically consists of two main processes: the forward process and the reverse process.

The forward process adds Gaussian noise to the data in a series of small steps, gradually transforming the data into pure noise as a Markov chain. Let $x_0$ represent the original data, $x_t$ the data at step $t$, $\mathcal{N}$ a Gaussian distribution, $\beta_t$ the variance of the noise added at step $t$, and I the identity matrix, which is a square matrix with ones on the diagonal and zeros elsewhere. The forward process can be described as a sequence of transitions, where the data at each step $t$, given the previous step $t-1$, is scaled down by $\sqrt{1-\beta_t}$ to keep the variance at 1 at every step $t$. The probability distribution is defined as:

$$q(x_t|x_{t-1}) = \mathcal{N}\big(x_t, \sqrt{1-\beta_t}x_{t-1}, \beta_t\mathrm{I}\big). \tag{2}$$

The reverse diffusion process aims to learn the transition probabilities to reverse the added noise and recover the original distribution within the data. This reverse process, parameterized by the mean $\mu_\theta$ and the variances $\Sigma_\theta$, can also be represented as a Markov chain. The probability distribution at each step $t-1$, given step $t$, is defined as:

$$p_\theta(x_{t-1}|x_t) = \mathcal{N}\big(x_{t-1}, \mu_\theta(x_t, t), \Sigma_\theta(x_t, t)\big). \tag{3}$$

The training objective of diffusion models is to minimise the difference between the forward and reverse processes, represented as a variational lower bound. This optimises the parameters $\theta$ of the neural networks and ensures that the learned reverse process can effectively denoise the data and recover the original distribution (Sohl-Dickstein et al., 2015; Ho et al., 2020).

3)    CHALLENGES

Training generative models can be challenging due to several factors. One major challenge in GANs is the stability of the training process, where one network becomes too strong and overpowers the other (Wiatrak et al., 2019). For instance, GANs can suffer from issues such as mode collapse, where the generator finds specific patterns that can fool the discriminator, leading to a lack of variety in the generated samples (Kushwaha & Nandi, 2020). Diffusion models, while generally more stable, require extensive computational resources and careful tuning of the noise addition and the reverse diffusion process. Despite these challenges, advancements in training techniques and model architectures have significantly improved the performance of such generative models (Bengesi et al., 2024).

Evaluating generative models also presents significant challenges. Different from typical supervised predictive models with ground truth for verification, no single metric can comprehensively assess the quality and diversity of the generated samples (Xu et al., 2018). There are many choices in evaluation metrics, each with strengths and weaknesses and providing insights into different aspects of model performance.

*f.*    *Related Studies*

Limited research currently focuses directly on this specific use case. However, a few studies have explored the application of generative models in meteorology and the use of synthetic data to improve predictive models in related fields.

Precipitation nowcasting is a domain that commonly demonstrates using generative models to generate realistic synthetic radar images and precipitation patterns (Asperti et al., 2023; Ravuri et al., 2021). Recent research also illustrates how generative models can be used in climate model downscaling, which enhances the spatial resolution of weather variables from coarse climate model outputs (Leinonen et al., 2020; Rampal et al., 2024). By generating high-resolution data, generative models can significantly improve the granularity and accuracy of climate projections, making them more useful for local-scale impact assessments.

Recent research by Miralles (2023) discusses the application of GANs to create synthetic datasets for studying the frequency and intensity of extreme weather events. On the other hand, a more applicable study by Chatterjee and Byun (2023) created synthetic time-series of weather parameters with GANs. By combining the synthetic data with original data, they significantly improved the accuracy of their ML-based predictive model on electric vehicle demand prediction, demonstrating the effectiveness of synthetic data in improving data-driven predictive models.

*g.*    *Research Objectives*

The primary objective of this research is to explore and validate the potential of generative models in producing realistic synthetic wind field data for the UK. This study aims to address the challenges posed by the rarity of extreme windstorms and the limited availability of reanalysis data, which impede comprehensive risk assessment and catastrophe modelling efforts in the insurance industry.

To achieve this, the research will employ hourly reanalysis data from the ERA5 dataset, from 1940 to 2022, provided by the ECMWF. The study will develop and train various generative models, including standard GAN, Wasserstein GAN with gradient penalty (WGAN-GP), and U-net diffusion models, to generate high-quality wind maps that accurately reflect the spatial characteristics and frequency of different wind patterns in the UK. The performance of these generative models will be rigorously evaluated using quantitative metrics, with some computed in a reduced dimensionality space using PCA.

Statistical analyses will be conducted to verify the fidelity of the generated synthetic data in replicating the distribution of wind speeds and patterns observed in the ERA5 dataset.

Furthermore, this research will discuss the potential integration of the synthetic wind field data generated by these models into existing catastrophe models used by the insurance industry. This approach could enhance the robustness and accuracy of these models, providing a more reliable basis for risk assessment and financial planning. By achieving these objectives, the study seeks to deliver valuable tools for the insurance industry and emergency management agencies, contributing to a better understanding and preparedness for these extreme windstorm events.

## 2. Data

*a. Data Description*

### 1) SOURCE AND COVERAGE

The dataset utilised in this study is the ERA5 reanalysis dataset produced by the European Centre for Medium-Range Weather Forecasts (ECMWF). ERA5 provides hourly estimates of a large number of atmospheric, oceanic, and land-surface variables (Hersbach et al., 2020). The ERA5 data undergo extensive post-processing and quality control, ensuring its accuracy and consistency for direct use in modelling (Campos et al., 2022).

Covering the period from 1940 to 2022, the dataset offers a high spatial resolution of 0.25° × 0.25°, which is sufficient for synoptic-scale wind pattern analysis. The geographical domain covered in this study is focused on the UK, with coordinates ranging from 49°N to 59°N in latitude and 8°W to 2°E in longitude, as shown in Fig. 2. This specific domain forms a data size of 40 × 40 and ensures that the dataset encompasses all relevant regions that could be affected by windstorms impacting the UK.

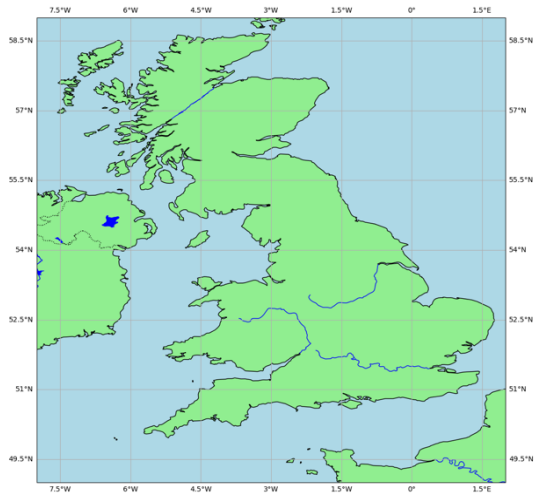

Figure 2. The spatial coverage of the ERA5 dataset used in this study.

### 2) VARIABLES USED

This study focuses primarily on the 10-metre hourly-averaged wind speed data from the ERA5 dataset, which is crucial for understanding wind patterns and their impact on the surface. The choice of the 10-metre wind speed is justified by its direct applicability to surface-level impacts, which are of primary concern in risk assessment and mitigation.

In addition to the 10-metre wind speed, the 10-metre 3-second wind gust data were also considered. Wind gusts, which represent short-term bursts of high wind speed, are particularly relevant for windstorm damage and risk assessment due to their potential to cause significant destruction (Gliksman et al., 2023). Initially, the models were trained and tuned using the wind speed data and later applied to the wind gust data. However, it was observed that the models performed poorly in extreme case scenarios when trained with wind gust data, sometimes producing noisy outputs. This indicated a sensitivity of the models to the input datasets and parameters.

Given the time constraints of this project, extensive tuning of the models specifically for the wind gust data was not feasible. Therefore, while wind gusts were tested and found to be a potential area for further study, the primary focus remained on 10-metre wind speed for this study to ensure the reliability of the models. Examples of the model performance using wind gust data, demonstrating the observed issues, will be provided in the Appendix.

3)    DATA PREPROCESSING

The 10-metre wind speed data were normalised to a range of $[0, 1]$ using global minimum and maximum values across the entire domain and timeframe. This transformation helps stabilise and accelerate the training process by ensuring numerical stability and enhancing the ability of the model to generalise (Ahmad & Aziz, 2019). When input features have widely varying scales, the optimisation algorithms can struggle with learning, often resulting in slower convergence and suboptimal performance. By scaling the input data uniformly, we mitigate issues related to varying scales of input features, leading to more efficient training and reducing the risk of overfitting.

4)    SPATIAL RESOLUTION CONSTRAINTS

One of the notable limitations of the ERA5 dataset is the spatial resolution constraints, specifically its handling of surface friction. Surface friction is crucial in influencing near-surface wind speed, particularly in regions with complex topography, such as coastal and mountainous areas (Laurila et al., 2021). The ERA5 dataset, despite its relatively high resolution, often fails to represent surface friction in these regions accurately. These inaccuracies stem from the coarse spatial resolution that cannot fully capture the fine-scale variations in land surface characteristics. High-resolution local variations, such as those caused by topography, land-sea interactions, and urban landscapes, are often smoothed out (Caton Harrison et al., 2022). This can result in underrepresenting extreme wind events and local wind maxima.

For instance, regions with significant elevation changes or varying land cover types can exhibit highly localised wind patterns that ERA5 may not resolve effectively. Studies have shown that ERA5 often underestimates wind speeds in high-elevation areas due to its inability to account for the ridge-acceleration effect and the Venturi effect, where wind speeds increase as air flows over mountain ridges or through narrow valleys (Potisomporn et al., 2023). Similarly, the transition between land and sea in coastal regions creates a complex interaction of thermal gradients and surface roughness, leading to higher biases and errors in wind speed estimates compared to flat, homogeneous terrains. A recent study by Dullaart et al. (2020) highlights that the ERA5 resolution still poses limitations for accurately modelling small-scale features, especially in tropical cyclone scenarios.

In this study, these surface friction inaccuracies are particularly relevant for regions like Scotland, where the highland areas could result in underestimated wind speeds. A land cover map of the UK (Fig. 3) highlights these areas of complex and inhomogeneous terrain, which are prone to such inaccuracies. While ERA5 provides valuable data for large-scale atmospheric patterns, caution must be exercised when analysing localised wind patterns.

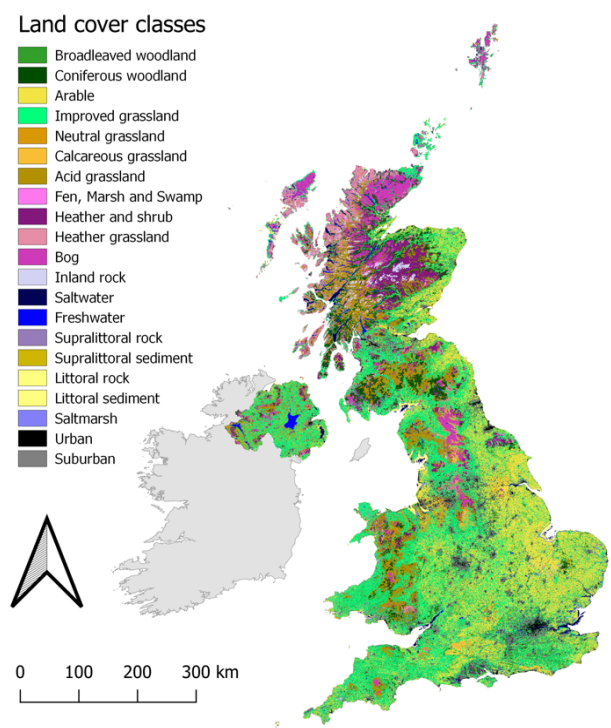

Figure 3. Land Cover Map 2021 of the UK (Marston et al., 2023). Regions like Scotland show varying land cover types, leading to higher biases in wind speed estimations.

## 3.    Methodology

### a.    Machine Learning Pipelines

The machine learning pipeline for this study is designed to systematically process, train, and evaluate the generative models for wind field simulation. The workflow diagram below (Fig. 4) illustrates the key stages of this pipeline. The process begins with data pre-processing, and the planning phase involves selecting appropriate variations of generative models and designing their architectures in detail, where the normalised data serves as training data for these models.

Following these preparatory steps, the models are trained, which involves iterative training to optimise model performance. Based on evaluation results, including visual inspection and quantitative metrics that assess the quality and diversity of the generated outputs, we adjust model parameters and architectures such as depth of the networks, input size of the latent space, and learning rates in the model tuning phase. After tuning, the models are trained again with the adjusted parameters. This iterative cycle of training and tuning continues until the model evaluation phase indicates satisfactory results. The final step in this project is to identify and select the best-performing version of the model for analysis and comparison between model variations.

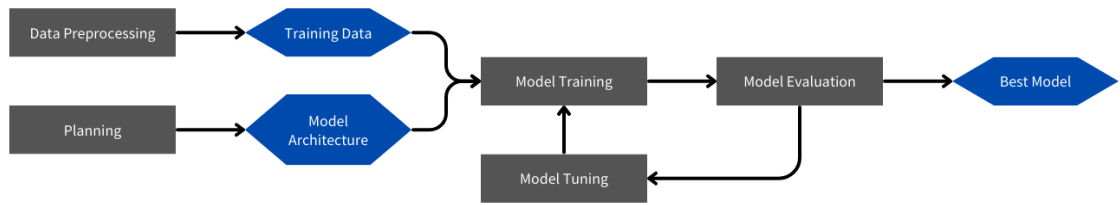

Figure 4. Overview of the machine learning pipeline for model development. Grey rectangles represent process stages, blue hexagons indicate objects, and arrows depict the flow of the pipeline.

### b.    Model Variations Explored

In this study, various generative models were initially trained and evaluated to identify the best-performing models for realistic wind field simulation. The models explored include standard GAN, WGAN, WGAN-GP, mean and covariance feature matching GAN (McGAN), variational autoencoder GAN (VAE-GAN), diffusion GAN, and U-net diffusion model. Over 300 model versions were trained and assessed, including various architectural variations, parameter configurations, and training processes. It has been found that the models are very sensitive to any change in the layers and parameters, usually leading to model instability.

The evaluation process relied heavily on qualitative assessments, such as visual inspections of the generated samples and monitoring of training loss. Visual inspection was crucial in identifying models that failed to produce realistic and diverse outputs, while training loss monitoring helped determine if the models were learning effectively. Models showing apparent underperformance or unchanging loss values were subjected to further tuning and adjustments. This iterative process continued until it was evident that no further improvements could be achieved. Examples of outputs from these less successful models, which were eventually not chosen for detailed analysis, are illustrated below for reference. These models typically encounter issues such as mode collapse (Fig. 5), significant underestimation of wind speeds (Fig. 6), and inability to resolve the noises added in the diffusion process (Fig. 7).

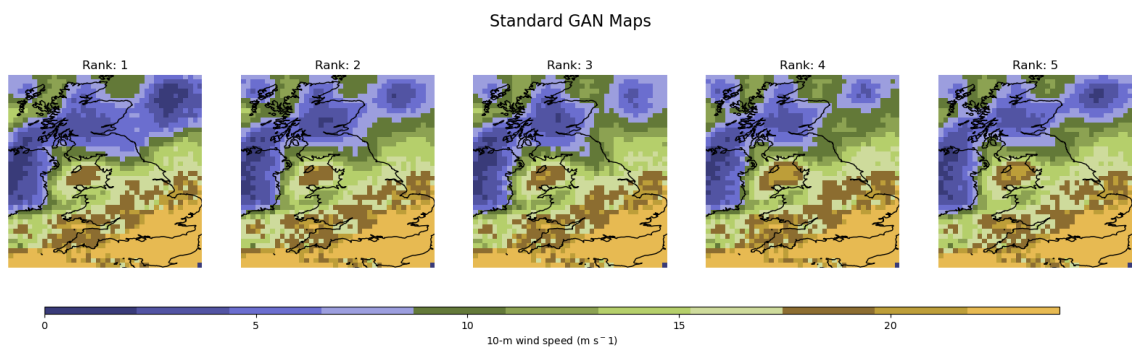

Figure 5. The wind speed maps with the greatest storm severity index from a standard GAN with no techniques implemented to address a mode collapse. The similar patterns across the maps indicate a significant model collapse, where the GAN consistently generates nearly identical outputs instead of capturing the variability expected in different scenarios.

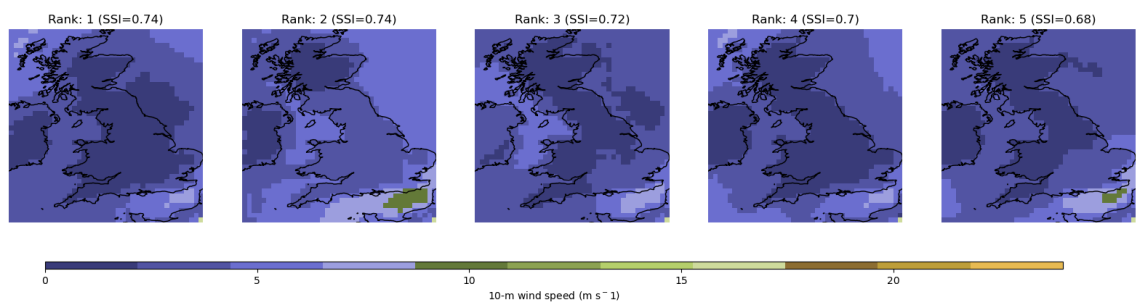

Figure 6. The wind speed maps with the greatest storm severity index from a diffusion model, with blue colours indicating wind speeds below 10 m s$^{-1}$. Despite a slight difference in the wind patterns among the maps, wind speeds are significantly underestimated, suggesting the inability of the model to replicate the magnitude of the wind field data.

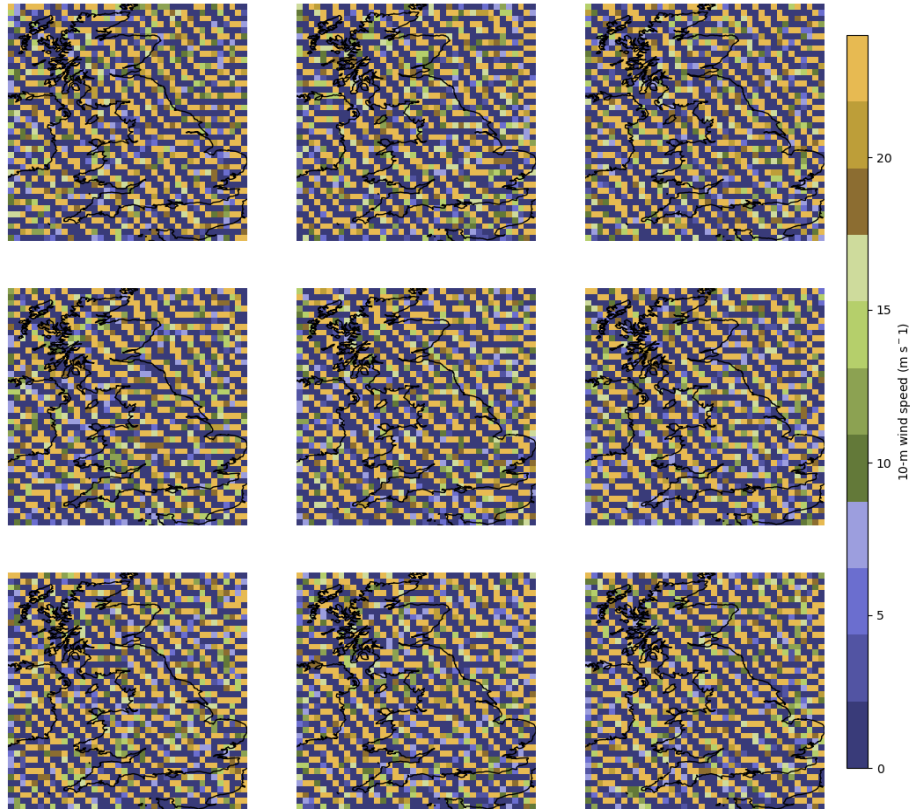

Figure 7. Random examples of wind speed maps generated from a diffusion GAN. The pixelated patterns illustrate that the model failed to resolve the noises added in the diffusion process.

When qualitative approaches could not clearly determine underperformance, quantitative metrics were used to evaluate the image quality and diversity more rigorously. Among these models evaluated, three were selected for detailed analysis due to their superior quantitative performance, including standard GAN, WGAN-GP, and U-net diffusion model. These models demonstrated better skill at producing realistic wind field simulations, particularly during extreme wind events, with significant diversity in the generated patterns. The detailed architectural and training differences are discussed in the following sections.

*c.    Selected Models for Detailed Analysis*

1)    STANDARD GENERATIVE ADVERSARIAL NETWORK (GAN)

The architecture of the standard Generative Adversarial Network (GAN) consists of two neural networks: the generator and the discriminator. The generator architecture is designed to transform a random noise vector into a synthetic wind speed map, while the discriminator architecture is tasked with distinguishing between real and generated wind speed maps.

The generator network begins with a low-dimensional noise vector with a dimensionality of 512 in the latent space, which is progressively transformed into a higher-dimensional output. This transformation involves a series of four transposed convolutional layers with a decreasing number of filters and increasing filter sizes, which up-sample the input to produce detailed and structured outputs. A transposed convolutional layer works by expanding each pixel in the input into a larger area in the output, which increases the spatial dimensions of the input and outputs the generated wind speed map with dimensions $40 \times 40 \times 1$.

To stabilise and accelerate the training process, batch normalisation is applied after each transposed convolutional layer. Batch normalisation normalises the inputs of each layer to have a mean of zero and a variance of one across the training batch. This process mitigates the issue of internal covariate shift, where the distribution of inputs to a layer changes during training, causing the network to struggle with convergence (Ioffe & Szegedy, 2015; Santurkar et al., 2018). A leaky rectified linear unit (LeakyReLU) activation function is followed by each batch normalisation layer. LeakyReLU is a variation of the rectified linear unit that allows a small, non-zero gradient when the unit is inactive due to a negative input. This helps prevent the "dying ReLU" problem, where neurons can become inactive and stop learning (Liu, 2021). This activation function ensures that there is a small gradient for negative inputs, allowing the network to continue updating and learning even when the input is negative.

On the other hand, the discriminator takes a $40 \times 40 \times 1$ wind speed map as input and passes it through a series of four convolutional layers with increasing filters and decreasing filter sizes. The convolutional layers down-sample the input by sliding the filter over the input to produce a feature map that highlights specific patterns. Each convolutional layer is followed by a LeakyReLU activation function, except for the final layer that employs the sigmoid activation function. This output represents the probability that the input data is real, making the sigmoid function as a binary classification that distinguishes between real and generated data.

The training of the standard GAN involves a two-step process where the discriminator and generator are trained alternatively. In each iteration, the discriminator is first trained on a batch of 64 real and generated images each, minimising the binary cross-entropy loss (Eq. 1). This helps the discriminator learn to differentiate between real and fake images effectively. The generator is then trained on a batch of 128 generated images, aiming to maximise the probability of the discriminator misclassifying the generated images as real. The training process was run for a total of 10,000 epochs, allowing the model to iteratively improve its performance.

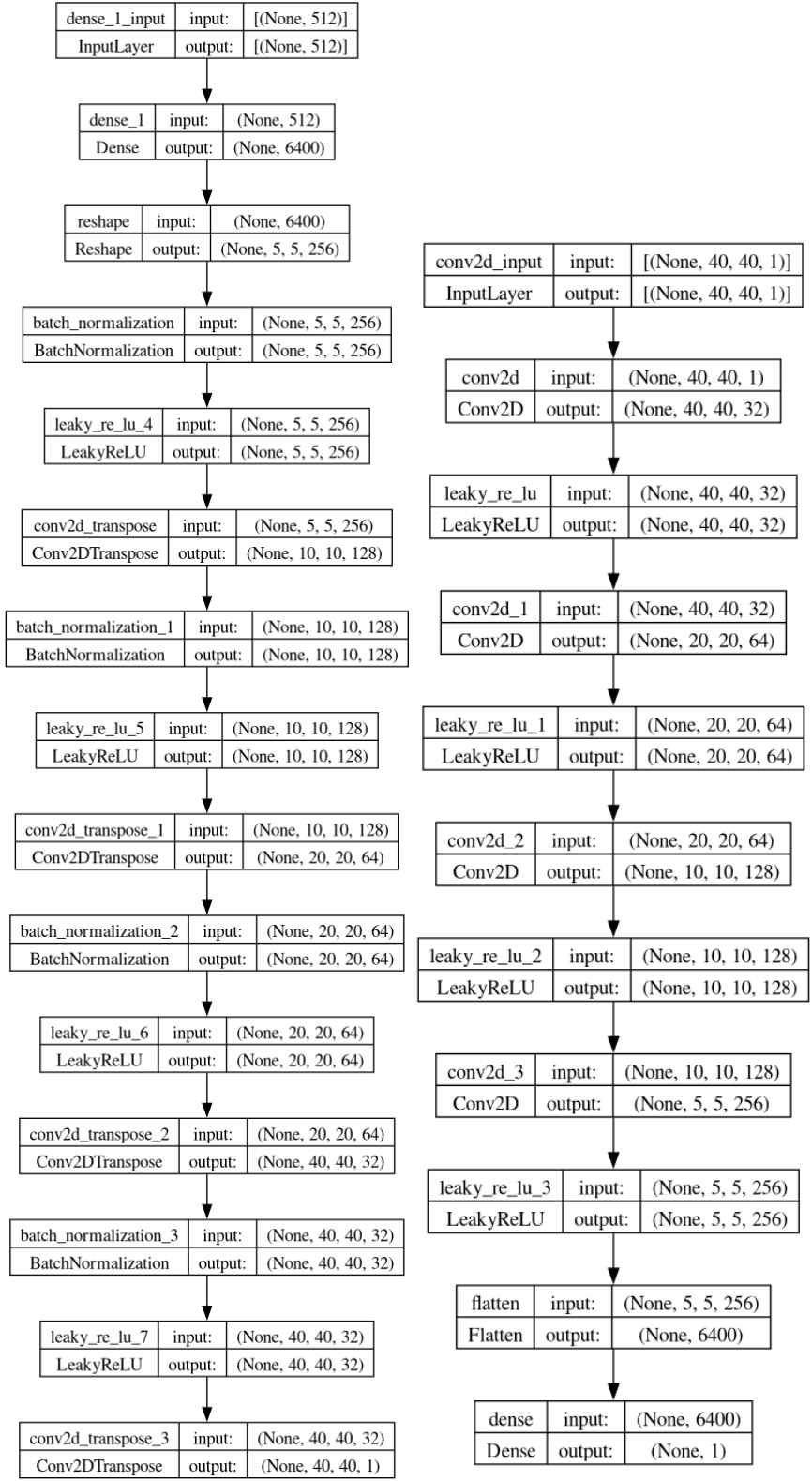

Figure 8. The architecture of the standard GAN shows the generator (left) and discriminator (right). The generator transforms a random noise vector with a dimensionality of 512 into a $40 \times 40 \times 1$ synthetic wind speed map through a series of transposed convolutional layers. The discriminator processes a $40 \times 40 \times 1$ wind speed map through convolutional layers to determine whether the input is real or generated.

## 2) WASSERSTEIN GAN – GRADIENT PENALTY (WGAN-GP)

The Wasserstein GAN with gradient penalty (WGAN-GP) is an advanced version of the standard GAN designed to improve training stability and prevent mode collapse. Like the standard GAN, the WGAN-GP consists of a generator and a discriminator, but it incorporates several key modifications.

The generator in the WGAN-GP shares a similar structure to the standard GAN. A unique addition in the WGAN-GP generator is the Gaussian noise layer before the first transposed convolutional layer, which introduces noise to the intermediate states to help regularise the model and prevent overfitting (You et al., 2019). The discriminator in the WGAN-GP, like in the standard GAN, processes the $40 \times 40 \times 1$ wind speed map through four convolutional layers with LeakyReLU activation functions. However, the final layer flattens the tensor and passes it through a dense layer that outputs a single scalar value, serving as a critic score. The removal of the sigmoid activation allows the use of a Wasserstein loss function, which measures the difference between the real and generated data distributions.

Unlike the standard GAN loss, which can suffer from vanishing gradients and mode collapse, the Wasserstein loss provides a smoother and more meaningful gradient, leading to more stable training (Arjovsky et al., 2017). The Wasserstein loss for the discriminator can be expressed as (Gulrajani et al., 2017):

$$L_{WL} = \mathbb{E}_{x \sim P_r}[D(x)] - \mathbb{E}_{z \sim P_z}\big[D\big(G(z)\big)\big]. \tag{4}$$

In this equation, $x$ represents real data samples, $z$ represents noise input to the generator, $P_r$ is the real data distribution, and $P_z$ is the noise distribution. The term $\mathbb{E}_{x \sim P_r}[D(x)]$ represents the expected value of the discriminator output for real data, and $\mathbb{E}_{x \sim P_r}[D(x)]$ represents the expected value of the discriminator output for generated data. The key concept here is that the Wasserstein loss measures the earth mover's distance (EMD) or the minimum cost of transporting mass to transform the generated data distribution into the real data distribution. In this context, a perfect critic score is -1, indicating the real data, and an imperfect score is 1, indicating the generated data.

To ensure the discriminator maintains the Lipschitz continuity constraint, a gradient penalty is introduced. Lipschitz continuity is a property that ensures the loss function has bounded gradients and does not change too rapidly while optimising (Thanh-Tung et al.,

2019). The gradient penalty enforces this by adding a regularisation term to the loss function, given by:

$$L_{GP} = \lambda \mathbb{E}_{\hat{x} \sim P_{\hat{x}}}[(\|\nabla_{\hat{x}} D(\hat{x})\|_2 - 1)^2]. \qquad (5)$$

In this equation, $\hat{x}$ represents interpolated samples between real and fake data, $\lambda$ is the gradient penalty coefficient, and the term $\|\nabla_{\hat{x}} D(\hat{x})\|_2$ is the L2 norm of the gradient of the discriminator output. The gradient norm is the magnitude of the gradient vector, and ensuring it is close to 1 helps maintain the Lipschitz constraint, providing smooth and reliable gradients for training the generator (Gouk et al., 2021).

In WGAN-GP, imbalance training ensures that the critic provides accurate gradients for the generator to learn from, meaning the discriminator is trained four times more frequently than each generator update. This imbalance helps the discriminator approximate the Wasserstein distance, leading to more stable training of the generator (Goodfellow et al., 2014). The training process was run for 5000 epochs, with batch normalisation and LeakyReLU activation functions retained to stabilise training and prevent neurons from becoming inactive. The addition of Gaussian noise in the generator layers also contributed to the ability of the model to generalise better and produce diverse outputs.

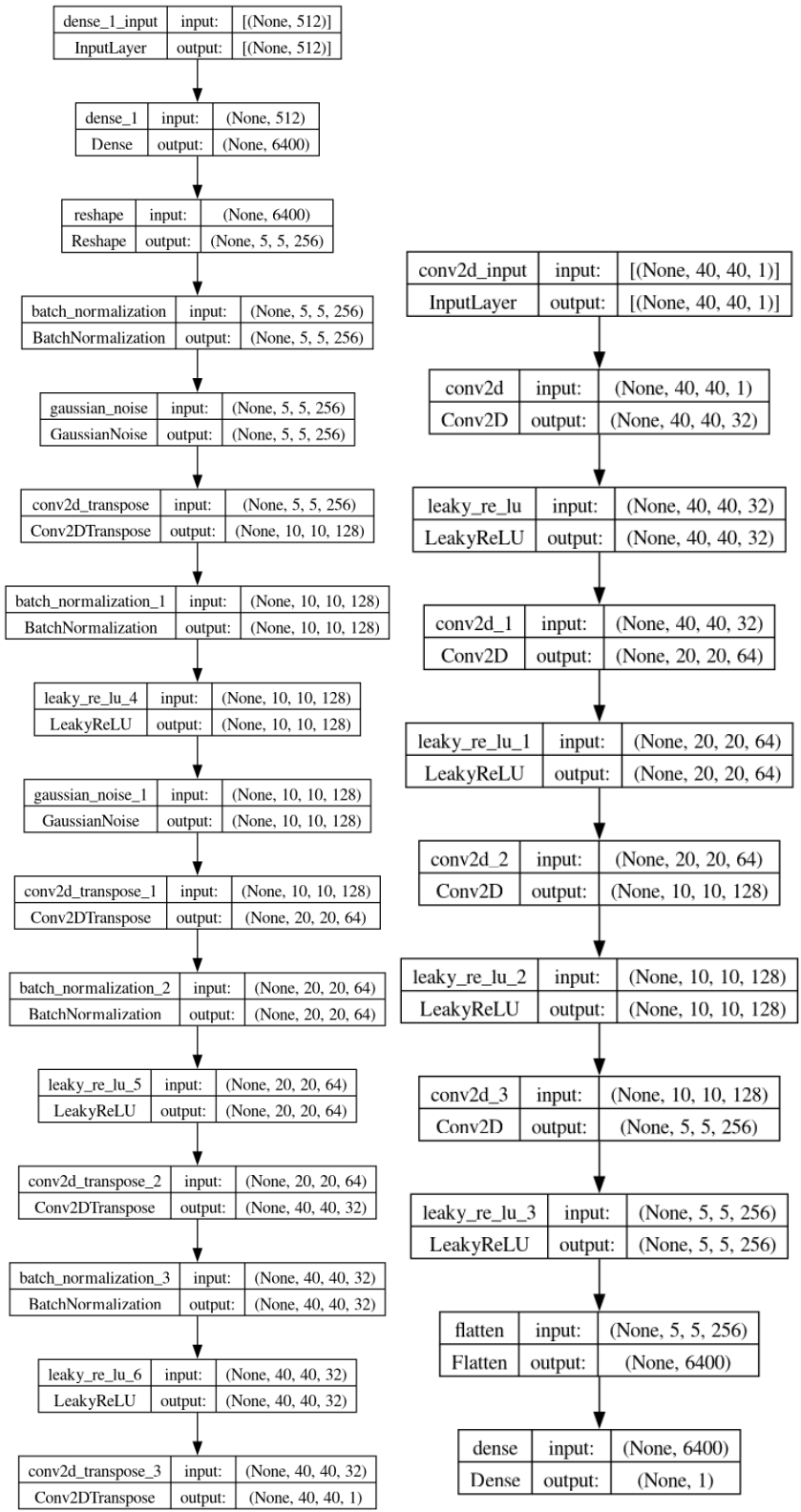

Figure 9. The architecture of the WGAN-GP shows the generator (left) and discriminator (right). The generator transforms a random noise vector with a dimensionality of 512 into a $40 \times 40 \times 1$ synthetic wind speed map through a series of transposed convolutional layers, with Gaussian noise layers added for regularisation. The discriminator processes a $40 \times 40 \times 1$ wind speed map through convolutional layers, outputting a scalar value used to calculate the Wasserstein distance for more stable training.

3)      U-NET DIFFUSION MODEL

The U-net diffusion model combined the architecture of a U-net with diffusion processes to generate realistic wind speed maps. The U-net architecture is characterised by a symmetrical structure consisting of an encoder and a decoder. The encoder compresses the input images into a lower-dimensional representation by progressively applying convolutional and pooling layers, while the decoder reconstructs the image back to its original dimensions using transposed convolutional layers. A U-net uses skip connection, where intermediate outputs from the encoder are concatenated with corresponding layers in the decode, aiming to retain spatial information lost during down-sampling and improve the quality of the generated images.

The generator of the U-net diffusion model begins with an input layer that accepts an image of dimensions $40 \times 40 \times 1$ and a timestep of the diffusion process. The input image is processed through a series of convolutional layers, each followed by batch normalisation and ReLU activation functions. The architecture is designed to progressively down-sample and then up-sample the spatial dimensions, creating a U-shaped structure.

Custom diffusion blacks are incorporated at multiple scales within this architecture, each consisting of convolutional layers and dense layers that integrate timestep information of the diffusion process, allowing the model to capture changes in wind speed data over different timesteps. Intermediate outputs from down-sampling layers are concatenated with corresponding up-sampling layers during the reconstruction phase, helping to retain spatial information at various scales.

The training process of the U-net diffusion model is designed to simulate the diffusion process through a sequence of timesteps. A forward noise function is used to add noise to the input images based on a series of timesteps, generating noisy images for each input image to represent different stages of the diffusion process. The model is trained over 250 epochs with a learning rate of 0.0008. During training, the model learns to denoise the images step by step by reducing the mean absolute error between the predicted and actual images.

To ensure effective training, the model performance is monitored by tracking the loss. Early stopping is triggered if the loss does not improve for 20 consecutive epochs, preventing overfitting. If the model does not show improvement, the learning rate is reduced by a factor of 0.75 to enhance convergence.

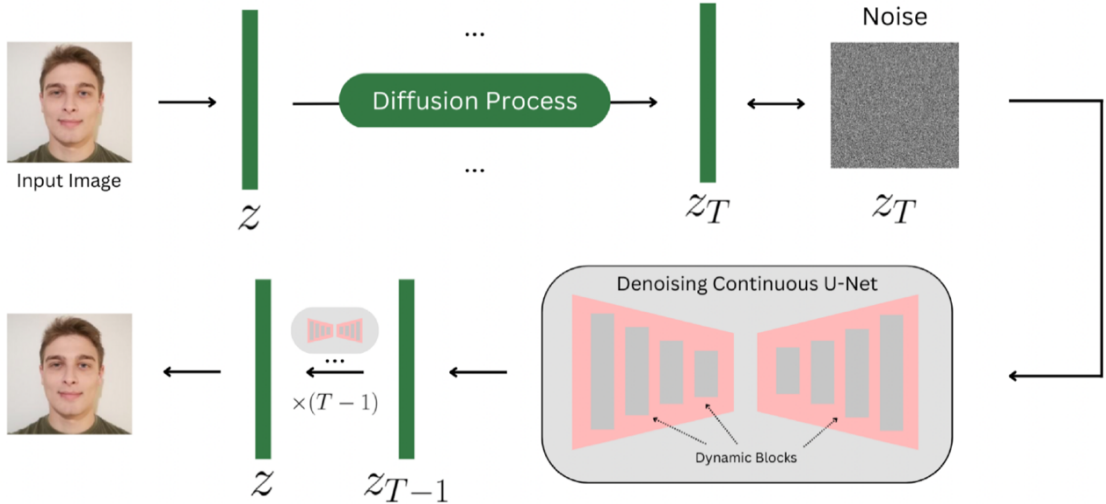

Figure 10. A brief architecture of the U-net diffusion model (Ordoñez et al., 2024). The model uses a series of convolutional layers in custom blocks to transform noisy images step by step into a $40 \times 40 \times 1$ synthetic wind speed map. The U-net involves skip connections between down-sampling and up-sampling layers to retain spatial information and improve the quality of generated outputs.

### d.    Evaluation Metrics

To quantitatively evaluate the performance of the generative models, we employed several metrics, each providing different insights into the quality, diversity, spatial characteristics, and statistical distribution of the generated wind speed maps. The following first introduces the key concepts underlying these metrics and the sampling of the datasets, followed by an overview of the metrics employed in this study and the rationale for their inclusion.

### 1)    INTRODUCTION TO KEY CONCEPTS

#### (i)    Storm Severity Index (SSI)

The storm severity index (SSI) provides a standardised method to quantify the severity of windstorms by accounting for both wind speed and the spatial extent of the storm (Klawa & Ulbrich, 2003; Leckebusch et al., 2008). It is particularly valuable in assessing the potential impact of different wind events in various locations. SSI is widely used in meteorological studies and the insurance industry to understand and manage the risks associated with windstorms. By evaluating their severity, SSI helps identify the most dangerous storms and the most vulnerable regions, which is crucial for interpreting the visual and quantitative evaluations of the models that follow.

The SSI of a wind event is calculated using the following formula (Klawa & Ulbrich, 2003):

$$SSI = \sum_{i=1}^{N} \left( \max \left( \frac{v_i}{v_{i,98}} - 1,0 \right) \right)^3, \qquad (6)$$

where $N$ is the number of grid cells in the domain, $v_i$ is the wind speed at grid cell $i$, and $v_{i,98}$ is the climatological 98th percentile wind speed at grid cell $i$.

This formula calculates SSI by summing the cubed normalised wind speed exceedances over the threshold $v_{i,98}$. By normalising the wind speeds against the 98th percentile, the formula accounts for regional variations in typical wind conditions, making the SSI a robust measure of storm severity and the spatial distribution of risks associated with windstorms. The choice of the 98th percentile as the threshold ensures that only significant wind speeds, which are more likely to cause damage, are considered in the calculation (Little et al., 2023).

In this study, the SSI is used in two ways. First, it is employed to determine extreme scenarios by identifying windstorms with the highest SSI values. This allows us to focus on the most severe events, analysing their spatial characteristics. Second, SSI serves as one of the quantitative evaluation metrics for assessing model performance to evaluate how accurately the models replicate the spatial distribution of severe wind events. By introducing SSI at the beginning of the section, we provide the necessary background to comprehend its relevance in the subsequent analyses.

*(ii)    Principal Component Analysis (PCA)*

Principal Component Analysis (PCA) is a statistical technique used to reduce the dimensionality of a dataset while retaining most of the variation present in the data (Jolliffe & Cadima, 2016). It transforms the original variables into a new set of uncorrelated variables called principal components (PC), which are ordered so that the first few retain most of the variation present in the original dataset. PCA is particularly valuable in meteorological studies for simplifying complex datasets, identifying scenarios, and highlighting significant features that contribute to variability (Jolliffe, 1990). In this study, PCA helps visualise the clusters of the wind speed data in a lower-dimensional latent space. We focus on simplifying the dataset for more efficient processing rather than ensuring that the modes of variability are physically realistic, as would typically be required in meteorological applications.

The PCA process in this study begins with scaling the data based on the grid data, which varies with latitude. This step accounts for the different spatial extents of the grid cells, ensuring that the contribution of each cell is proportional to its area, which is not true in a regular latitude-longitude grid. The scaling factor for each grid cell $i$ can be represented as:

$$Scaling\ Factor_i = \cos(\phi_i), \tag{7}$$

where $\phi_i$ is the latitude of the grid cell $i$. Subsequently, each dataset is standardised separately and locally for each grid cell, ensuring that the data is centred and scaled appropriately. The standardised value $z_{ij}$ for grid cell $i$ and sample $j$ is computed as:

$$z_{ij} = \frac{x_{ij} - \mu_i}{\sigma_i}, \tag{8}$$

where $z_{ij}$ is the original value, $\mu_i$ is the mean, and $\sigma_i$ is the standard deviation of the data in grid cell $i$.

After scaling and standardising the data, the standard PCA process is applied. This involves computing the covariance matrix of the standardised real data, extracting the eigenvalues and eigenvectors, and transforming the data into the new PC space (Wold et al., 1987). The generated datasets are then projected onto this same PC space defined by the real data using the eigenvectors.

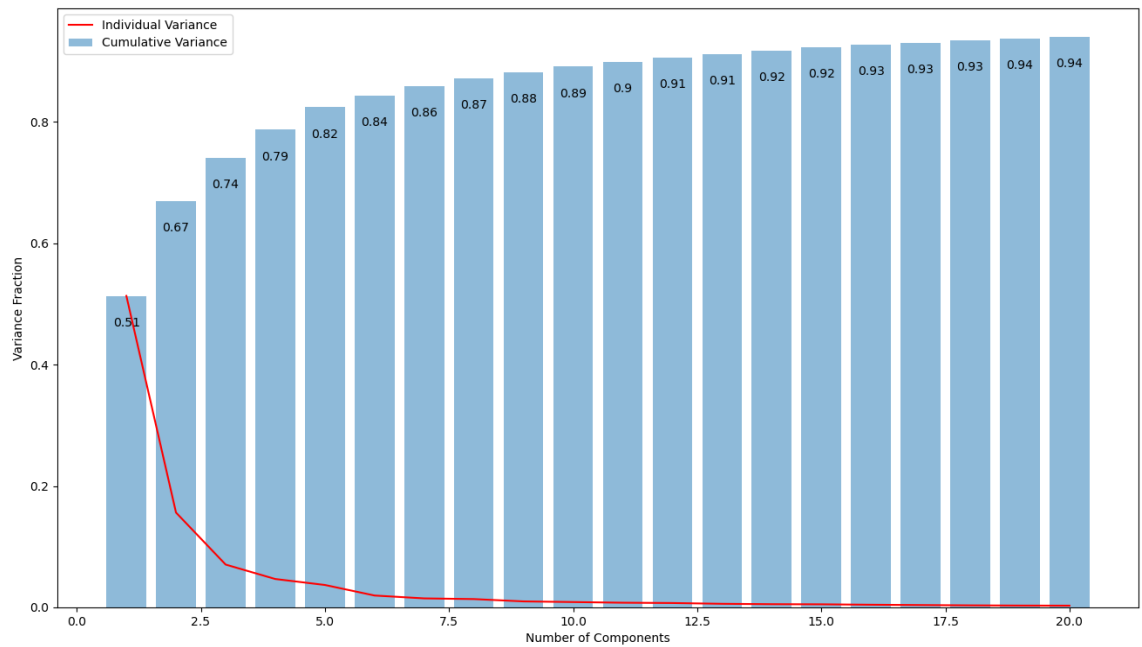

Figure 11. The plot shows the individual variance explained by each component, represented by the red line, and the cumulative variance, represented by the blue bars. The first 11 PCs explain more than 90% of the variance, indicating that the dimensionality of the dataset can be effectively reduced with minimal loss of information.

There is a difference in sample sizes between the ERA5 dataset and the generative models. The ERA5 dataset contains approximately 700,000 samples, whereas the generative models produced 200,000 samples each. This discrepancy is due to memory constraints in the hardware, which restricted the number of samples that could be generated from the models and influenced the model ability to generate the same variability and extremes of wind speed distributions as the ERA5 dataset.

Another important consideration is the nature of the samples generated by the models. The generative models produce independent and identically distributed (IID) samples, meaning each generated wind speed map can be considered as representing a distinct storm event with no temporal correlation between samples. In contrast, the ERA5 dataset includes temporal information, where the same storm event may be recorded at different hours, resulting in multiple instances of high SSI values for the same event. This complicates the comparison of extremes between the actual and generated data.

To address the above issue while comparing wind speed maps of the most damaging storms, a specific filtering process was applied to the ERA5 dataset by excluding events occurring within 72 hours of each other to ensure that only the most severe SSI case for each storm was considered. However, this filtering was not applied to other plots and evaluation metrics, where all available samples from the EAR5 dataset were used without additional temporal filtering.

One potential solution to achieve a more direct comparison between the ERA5 dataset and the generative models would be to randomly subsample the ERA5 data to match the sample size of the generated data. However, this approach was not applied in this study due to time constraints and some practical challenges. Subsampling could risk losing extreme events that are vital for an evaluation of the model performance on handling extremes and ensuring that the subsampled data remains representative of the full dataset is necessary.

Given these factors, although slightly lower winds might be expected for the tail-end scenarios, 200,000 samples (equivalent to 23 years of hourly data) are still sufficient to capture a wide range of scenarios and most extreme cases in the ERA5 dataset. The current approach still provides insights into how the generative models replicate the overall distribution and handle extreme wind events. However, future work could subsample techniques or other methods to enhance the comparison.

3) INTRODUCTION TO EVALUATION METRICS

*(i)    Fréchet Inception Distance (FID)*

The Fréchet inception distance (FID) is a common metric used to evaluate generative models, which compares the overall distributions of feature representations extracted from real and generated images using a pre-trained InceptionV3 model (Borji, 2022). The InceptionV3 model is a deep convolutional neural network (DCNN) designed for image recognition tasks and has been pre-trained on the ImageNet dataset, which contains millions of images across thousands of categories (Krizhevsky et al., 2012; Szegedy et al., 2016). Due to this extensive training, the InceptionV3 model is very good at recognising and classifying a wide variety of images.

However, wind speed maps are not included in the ImageNet dataset, which means that the feature extraction performed by the InceptionV3 model may not correspond directly to physical structures present in meteorological data. Wind speed maps generally have much lower entropy compared to the diverse set of images in ImageNet. As a result, the FID score primarily indicates the image-based visual and statistical similarity between the generated and real images rather than their physical accuracy.

To compute the FID score, both real and generated wind speed maps are passed through the InceptionV3 model to extract feature representations from one of the intermediate layers. The mean and covariance of these features are calculated for both real and generated images. The Fréchet distance between the two distributions is then computed using the following formula (Heusel et al., 2017):

$$FID = \left\| \mu_r - \mu_g \right\|^2 + Tr\left( \Sigma_r + \Sigma_g - 2\sqrt{\Sigma_r \Sigma_g} \right). \tag{9}$$

In this equation, $\mu_r$ and $\mu_g$ represent the mean feature representations of the real and generated data distributions respectively, while $\Sigma_r$ and $\Sigma_g$ are the corresponding covariance matrices. The term $\left\| \mu_r - \mu_g \right\|^2$ represents the squared difference between the mean vectors, aiming to capture the difference in average features between the real and generated datasets, while $Tr\left( \Sigma_r + \Sigma_g - 2\sqrt{\Sigma_r \Sigma_g} \right)$ is the trace of the sum of the covariance matrices, which considers the differences in the variability and correlation structure between two distributions. A lower FID score indicates that the generated images are more similar to real images in terms of their structural content and variety.

Another metric used to evaluate the performance of the generative models is the structural similarity index measure (SSIM) of the average SSI map, which measures the spatial accuracy of the generated wind speed patterns by comparing them to ERA5 dataset, specifically focusing on extreme scenarios. This metric evaluates the similarity between two images based on three aspects: luminance (mean intensity), contrast (variation in intensity), and structure (spatial arrangement of intensities).

We first calculate the SSI values for each sample and grid point in the generated and real wind speed maps and then create the average SSI map by averaging these values across all samples for each grid point. This map represents the typical intensity and distribution of severe wind speeds over the studied area and highlights regions that are more vulnerable to extreme wind events. The SSIM is then used to compare the average SSI maps of the generated and ERA5 data using the following formula (Wang et al., 2004):

$$SSIM(x, y) = \frac{\left(2\mu_x\mu_y + C_1\right)\left(2\sigma_{xy} + C_2\right)}{\left(\mu_x{}^2 + \mu_y{}^2 + C_1\right)\left(\sigma_x{}^2 + \sigma_y{}^2 + C_2\right)}. \tag{10}$$

In this equation, $x$ and $y$ represent patches from the real and generated average SSI maps respectively. The terms $\mu_x$ and $\mu_y$ are the mean values of these patches that capture the luminance, $\sigma_x{}^2$ and $\sigma_y{}^2$ are the variances of these patches that capture the contrast, and $\sigma_{xy}$ is the covariance between the two patches that reflects their structural similarity. The constants $C_1$ and $C_2$ are small values included to avoid computational error when the means or variances are close to zero.

This score ranges from -1 to 1, where 1 indicates perfect similarity, 0 indicates no correlation, and -1 indicates perfect negative correlation. A higher SSIM close suggests that the generated dataset closely resembles the ERA5 dataset in terms of spatial patterns and structural characteristics of the average SSI map. This approach focuses on assessing how well the model replicates the spatial characteristics of extreme wind events rather than the general spatial patterns. The performance can also be influenced by how the models handle extreme wind speeds.

*(iii)* *PCA-based Analysis*

To further evaluate the performance of the generative models, we applied PCA to reduce the dimensionality of the wind speed maps and analyse the distributional similarity between the ERA5 dataset and the models. The mean Kullback-Leibler (KL) divergence and the Earth mover's distance (EMD) were calculated on the first 25 PC dimensions, which explain approximately 95% of the variability in the EAR5 dataset. These metrics provide insights into how closely the generated data matches the real data in this reduced-dimensional space. Since the PCs have been sorted based on their importance, they reflect the amount of variability they capture. Therefore, no additional weighting is necessary to apply on the PCs while computing the metrics.

The mean KL divergence measures the difference between the probability distributions of the real and generated data in the PCA-reduced space (Shlens, 2014). It is calculated as:

$$KL(P \parallel Q) = \sum_{i=1}^{N} P(x_i) \log\left(\frac{P(x_i)}{Q(x_i)}\right), \tag{11}$$

where $x_i$ represents the data point of the PC, $N$ represents the total number of data points in the dataset, $P(x_i)$ and $Q(x_i)$ represent the probability distributions of the ERA5 and generated data respectively. KL divergence is particularly sensitive to small differences in the distributions and is useful for detecting finer discrepancies in how the two distributions diverge (Rekavandi et al., 2021). A lower mean KL divergence indicates that the generated data distribution closely matches the real data distribution, suggesting that the generative models effectively capture the details in the underlying variability of the ERA5 dataset.

The EMD, also known as the Wasserstein distance, measures the distance between two probability distributions by considering the minimal amount of work required to transform one distribution into another (Rubner et al., 2000). It can be calculated as:

$$EMD(P, Q) = \inf_{\gamma \in \Gamma(P,Q)} \mathbb{E}_{(x,y)\sim\gamma}[\|x - y\|], \tag{12}$$

where $x$ and $y$ represent individual data points in the PCA-reduced space from the ERA5 and generated datasets respectively. The term $\Gamma(P, Q)$ denotes the set of all possible joint probability distributions $\gamma(x, y)$ that have $P$ and $Q$ as their marginal distributions, while the expected value $\mathbb{E}_{(x,y)\sim\gamma}[\|x - y\|]$ represents the average distance between pairs of points $x$ from the ERA5 data and $y$ from the generated data, averaging over the joint distribution $\gamma$. The infimum represents the minimal value of the expected distance overall over possible joint distributions $\gamma$.

Unlike the KL divergence, which can become large for small differences in the distributions, EMD provides a more robust measure of similarity by focusing on the overall distance between the distributions, indicating broader and more global distributional differences. In the context of the PCA-reduced space, EMD measures the distance between the distributions of real and generated data across the first 25 PCs. Similar to the KL divergence, a lower EMD indicates a closer match between the distributions, reflecting the ability of the generative models to replicate the variability and distribution of the ERA5 dataset in general.

While the mean KL divergence detects detailed, fine-scale discrepancies, EMD is better suited for assessing global differences between distributions. By using both metrics, the evaluation can capture different aspects of the distributional similarity, identifying whether the models are consistently good across the whole data distribution.

*e.      Survey Design and Methodology*

An online survey was designed to gather subjective evaluations of the generated wind speed maps from a diverse group of participants within the university department with varying levels of expertise and experience in meteorology. The aim was to assess the perceived realism and physical plausibility of the maps, which might not be fully captured by objective metrics.

The survey involved presenting respondents with 20 wind speed maps of the UK. These maps were divided with 5 maps each from the ERA5 dataset and each of the three models. To ensure a comprehensive evaluation, the maps from each dataset were randomly selected based on their SSI values. The selection criteria included two maps from the top 25% of SSI values, two maps from the 25% to 75% range of SSI values, and one map from the bottom 25% of SSI values. These 20 maps were shuffled randomly and shown to the respondents in the same order to ensure consistency. Respondents were asked to rate the realism of each map on a Likert scale from 1 to 5, with 1 indicating "not realistic at all" and 5 indicating "very realistic". This approach allows for a quantitative assessment of how convincingly each model can replicate real-world wind patterns.

At the beginning of the survey, respondents were asked to categorise themselves into one of five proficiency levels, as listed in the pie chart (Fig. 12), which illustrates the distribution of the self-reported proficiency levels of the respondents. The diverse respondent profile ensures that the evaluation incorporates perspectives from various levels of expertise, which is crucial for obtaining a balanced evaluation.

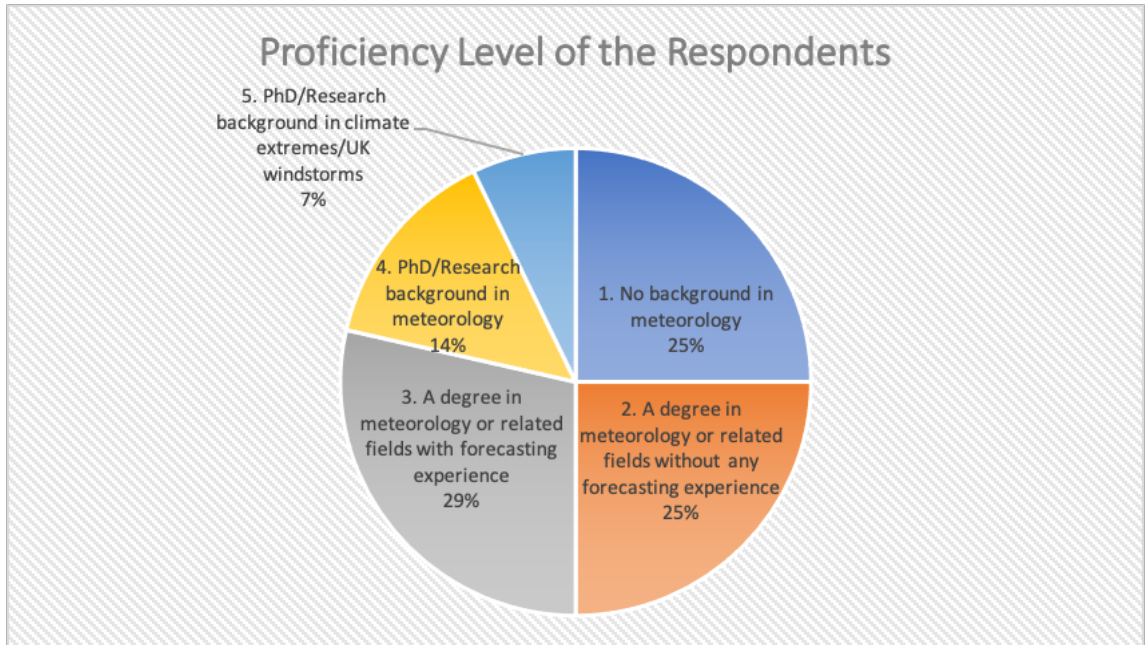

Figure 12. Distribution of the self-reported proficiency levels of the respondents in the human evaluation survey. The proficiency levels are categorised as follows: No background in meteorology (25%), a degree in meteorology or related fields without any forecasting experience (25%), a degree in meteorology or related fields with forecasting experience (29%), PhD/Research background in meteorology (14%), and PhD/Research background in climate extremes/UK windstorms (7%).

*f.    Training Environment*

1)    HARDWARE SPECIFICATION

The training of the models was performed on a high-performance computing setup equipped with four NVIDIA A100-SXM4-40GB graphics processing units (GPUs). Each GPU offers 40 GB of random-access memory (RAM), providing sufficient capacity for training ML models. The NVIDIA A100 is known for its exceptional performance and efficiency, with each unit capable of delivering up to 400 watts of power. The training datasets occupied 4.4 GB of storage, which was easily accommodated by the available GPU memory. The use of multiple GPUs enabled parallel processing, significantly reducing the training time and allowing large datasets and complex model architectures.

2)    SOFTWARE LIBRARIES

The training environment was set up on a Linux operating system with a Compute Unified Device Architecture (CUDA) version of 12.4, which facilitated GPU acceleration and optimised performance. Tensorflow version 2.15.0 was used as the primary deep learning framework, leveraging its comprehensive libraries and tools for building and training the generative models.

# 4. Results

*a.    Visual Evaluation of Generated Outputs*

1)    TYPICAL SCENARIOS

*(i)    Wind Speed Map Examples*

Fig. 13 shows nine random samples of 10-metre hourly wind speed maps from the ERA5 dataset and each of those generated by the three models introduced earlier. By analysing these maps, we aim to understand how well the models generally replicate the typical wind patterns observed in the ERA5 dataset.

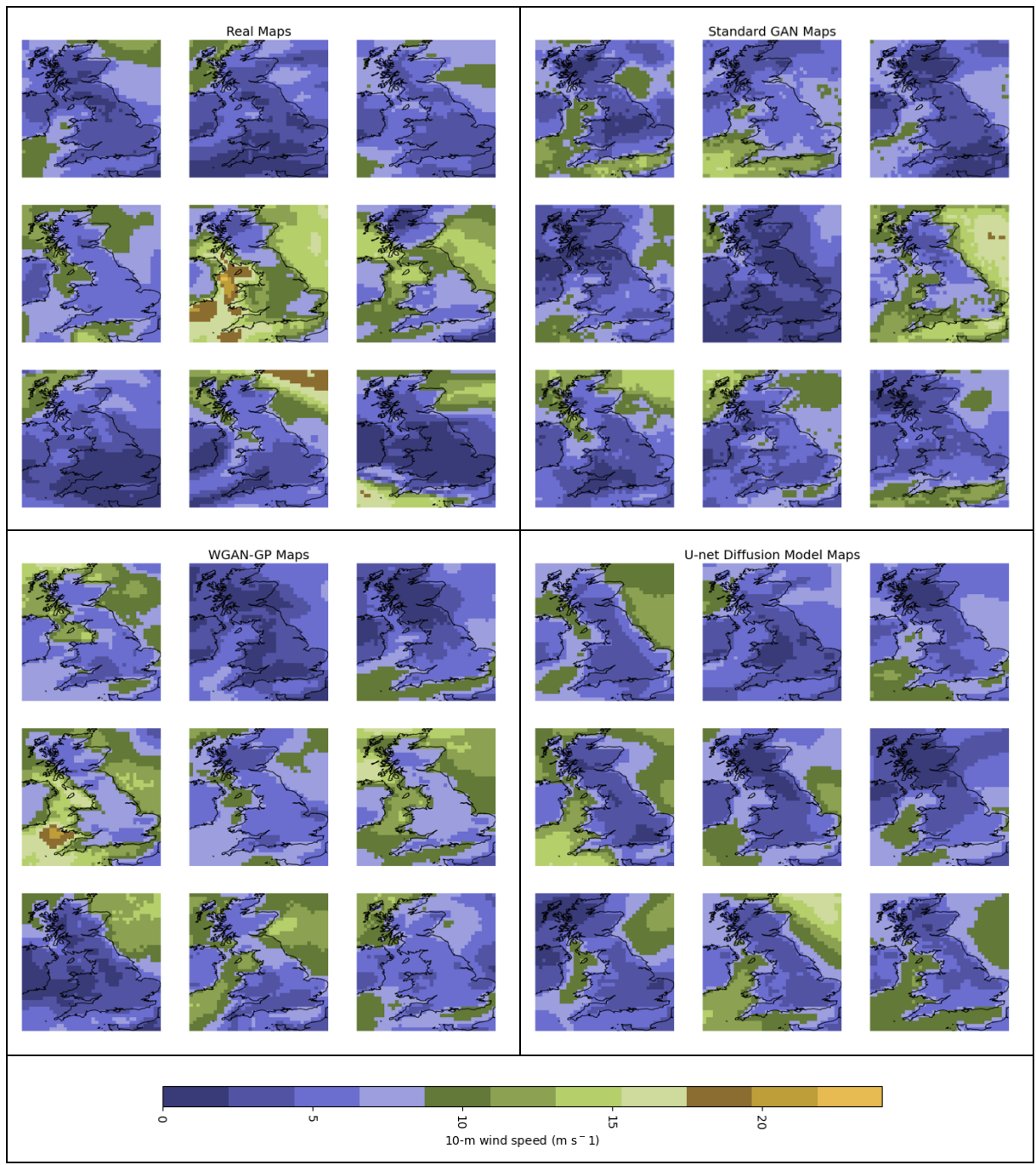

Figure 13. Comparison of typical wind speed maps. The plot shows nine random 10-metre wind speed maps in meters per second (m s⁻¹) from the ERA5 dataset (top left), standard GAN (top right), WGAN-GP (bottom left), and U-net diffusion model (bottom right).

The ERA5 wind speed maps (top left) exhibit certain consistent features with higher wind speeds over open waters (generally in dark green) and slightly higher wind speeds in coastal regions (generally in light blue). Blue colours indicating wind speeds lower than 10 m s⁻¹ are typically observed in inland regions, particularly southern England and Scotland. Similar general spatial characteristics can also be identified in the generated maps from all three models, indicating that these models are able to replicate the patterns and magnitude of the typical scenarios observed in the ERA5 dataset. However, these maps are sampled randomly, and they may not represent all possible scenarios. Therefore, these maps only serve as an overview of the results, and further quantitative evaluation is necessary.

Moreover, it is observed that some generated outputs from the standard GAN (top right) and WGAN-GP (bottom left) models show more noise, with a less smooth gradient, compared to the ERA5 maps and those generated by the U-net diffusion model (bottom right). This observation suggests that there are slight differences in the image quality of the generated outputs between different models. Moreover, due to the surface friction constraints in the ERA5 dataset, wind speeds over complex terrains such as the Scottish Highlands appear lower than expected, which should be considered when visually evaluating the model performance.

*(i)*      *Contour Plots on PC Space*

Fig. 14 illustrates the distribution of all samples on the first two PC dimensions, which contribute to around two-thirds of the variability. The red dotted lines in the plots represent the PC vectors, also known as the eigenvectors for these two components, which indicate the directions in which the variance in the data is maximised. These plots are based on the same sets of samples for the wind speed maps, where the ERA5 dataset has more samples than the others. By analysing these plots, we can assess how well the models replicate the key modes of variability observed in the ERA5 dataset.

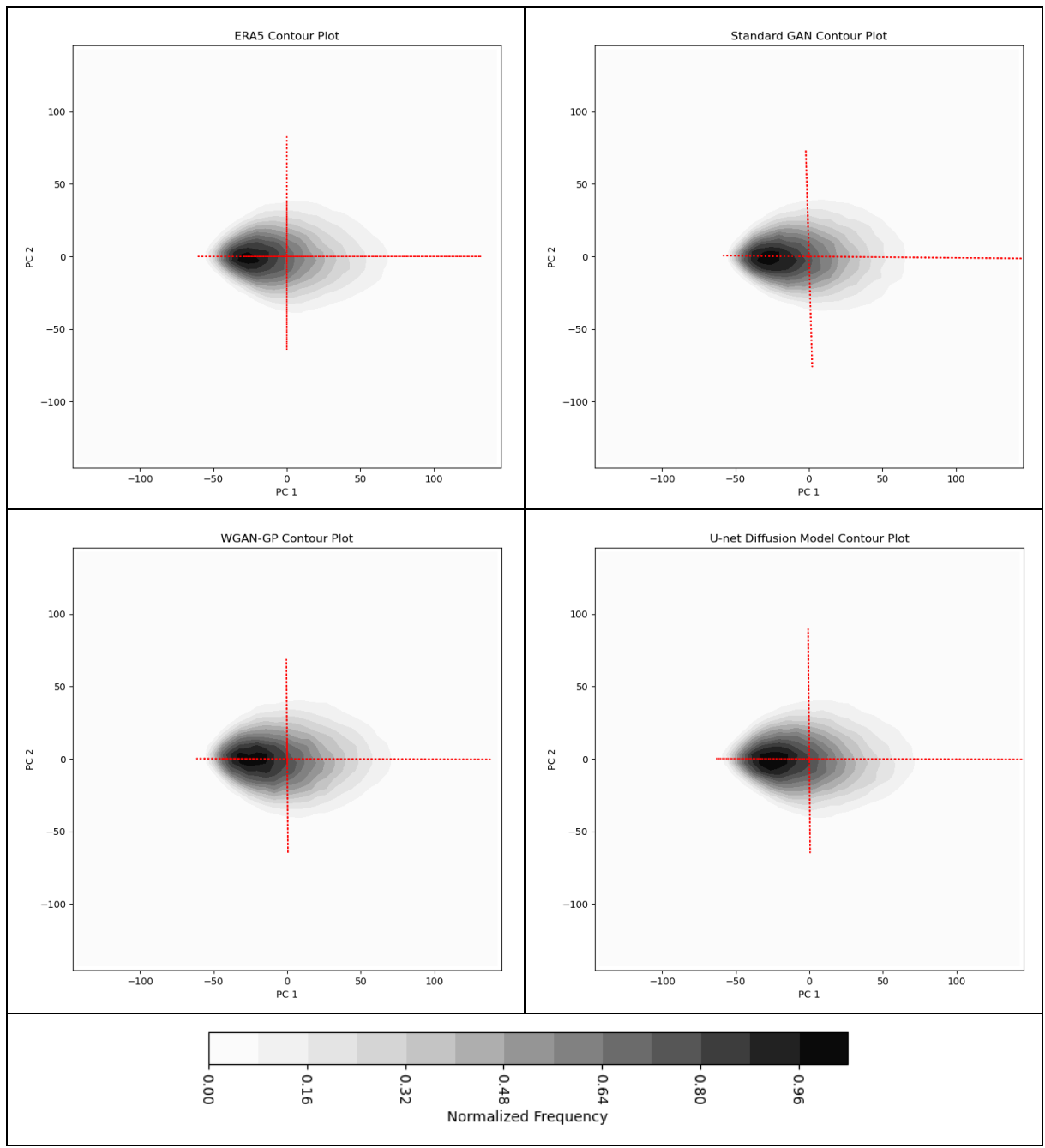

Figure 14. Contour plots of all samples on the PC1 (horizontal) and PC2 (vertical) dimensions for the ERA5 dataset (top left), standard GAN (top right), WGAN-GP (bottom left), and U-net diffusion model (bottom right). The colour bar indicates the normalised frequency of occurrences, while the red dotted lines represent the PC component vectors.

The contour plots for the ERA5 dataset and the three models on the first 2 PC dimensions show similar overall density and spread of the contours that is relatively concentrated around the origin. The contours extend along the direction of the PC1 vector and display a skewness towards the negative side on the PC1 dimension, while the PC2 vector is orthogonal to PC1. However, the orientation and length of the PC vectors vary slightly among the models.

For the standard GAN, the PC1 vector is observed to rotate slightly anti-clockwise compared to the ERA5 dataset and the other two models. This rotation indicates a shift in how the standard GAN interprets the primary mode of variability in the data. Additionally, the PC1 vectors for the standard GAN and WGAN-GP appear to be slightly shorter and positioned slightly downward compared to the ERA5 dataset and the U-net diffusion model, suggesting that they might be underrepresenting the primary mode of variability along the PC1 dimension.

On the other hand, the PC2 vectors for the standard GAN and the U-net diffusion model extend slightly further to the right than in the ERA5 dataset and the WGAN-GP, indicating a slight increase in the variability captured along the PC2 dimension. This extension suggests that they may be overemphasising the second mode, potentially leading to a slight imbalance in how the variance is distributed along the PC space. While the models can replicate the overall distribution of the primary variability, these differences in the orientation and length of the PC vectors across the models show variations in how each model captures the structure of the wind speed data.

2)  EXTREME SCENARIOS

*(i)    Wind Speed Map Examples*

In this section, we analyse the extreme windstorm scenarios by comparing the top 10 SSI cases over land from the ERA5 dataset (Fig. 15) with those generated by the three models (Fig. 16, 17 & 18). This comparison evaluates how well each model captures the intensity and spatial characteristics of extreme wind events.

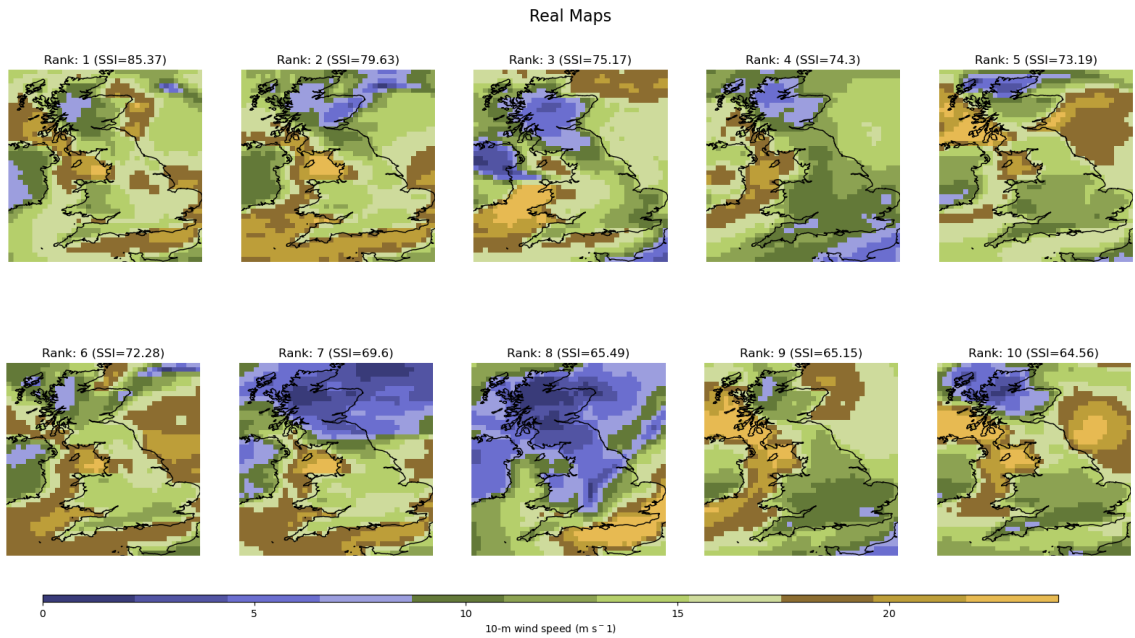

Figure 15. Top 10 SSI-ranked instances of extreme windstorm scenarios from the ERA5 dataset. The maps illustrate the 10-metre wind speed (in m s⁻¹) across the UK, showing areas with the highest wind speeds shaded in brown and yellow.

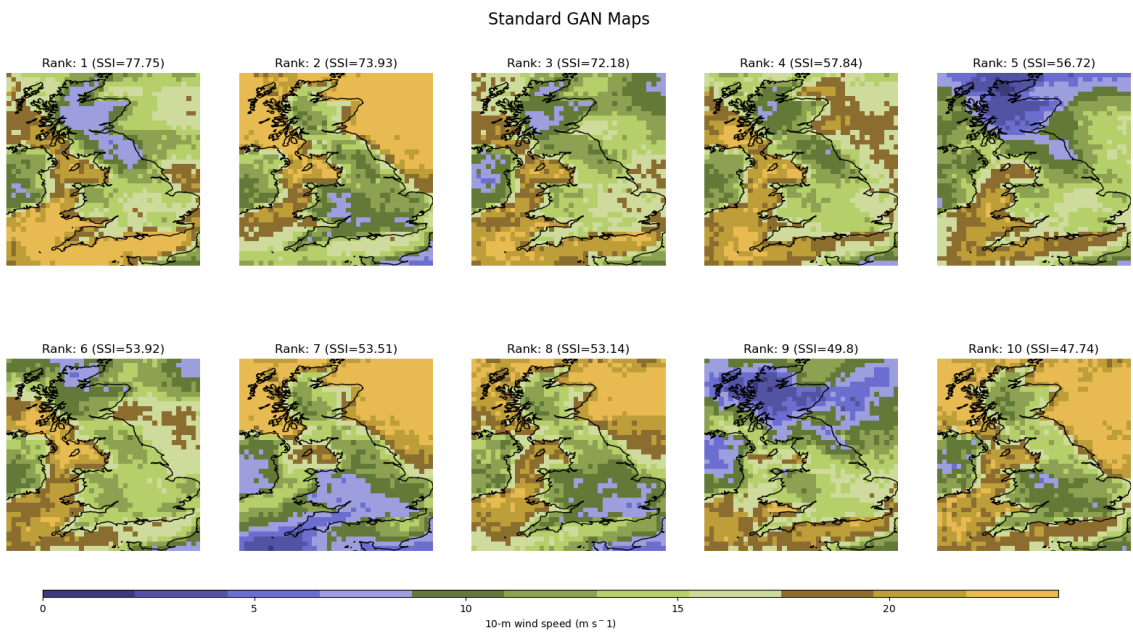

Figure 16. Top 10 SSI-ranked instances of extreme windstorm scenarios from the standard GAN. The maps illustrate the 10-metre wind speed (in m s⁻¹) across the UK, showing areas with the highest wind speeds shaded in brown and yellow.

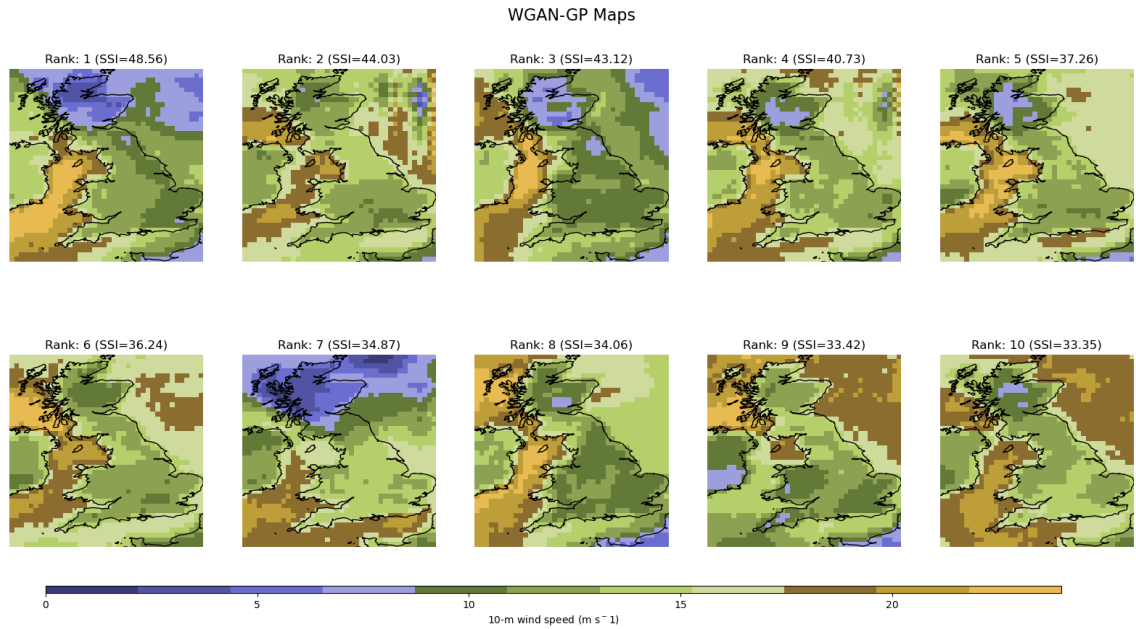

Figure 17. Top 10 SSI-ranked instances of extreme windstorm scenarios from the WGAN-GP. The maps illustrate the 10-metre wind speed (in m s$^{-1}$) across the UK, showing areas with the highest wind speeds shaded in brown and yellow.

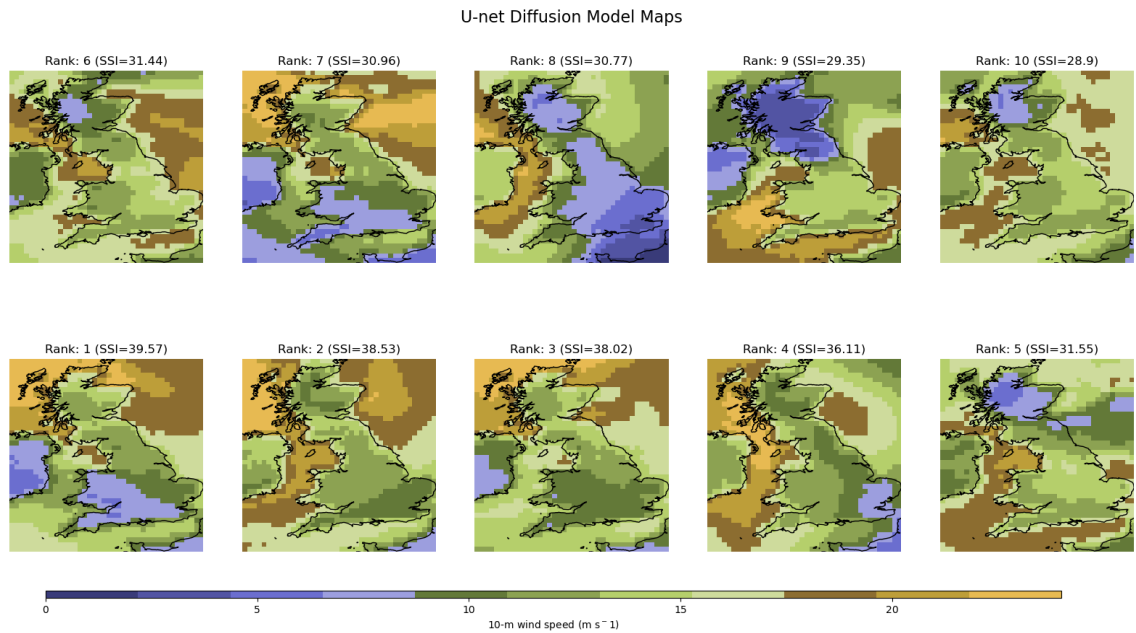

Figure 18. Top 10 SSI-ranked instances of extreme windstorm scenarios from the U-net diffusion model. The maps illustrate the 10-metre wind speed (in m s$^{-1}$) across the UK, showing areas with the highest wind speeds shaded in brown and yellow.

Among the top 10 SSI cases, the ERA5 dataset (Fig. 15) reveals spatial patterns similar to those of the typical scenarios but with higher variability in intensity and distribution. High wind speeds over 18 m s$^{-1}$ (in brown and yellow) are observed in some areas of open waters, such as the North Sea, the English Channel, and the Irish Sea. On the other hand, lower wind speeds are sometimes observed over inland regions such as south England, Scotland, and Ireland. While all three models successfully capture these spatial characteristics and identify the regions prone to extreme winds, several weaknesses were observed.

The standard GAN (Fig. 16), while successfully identifying the regions with higher wind speeds, often overextends the regions with extreme winds, resulting in slightly larger areas appearing in brown or yellow. Similar to its performance in typical scenarios, noisier maps with a less smooth gradient are also observed. Moreover, the 7th ranked map shows low wind speeds (in blue) near Southwest England and the Southwest Approaches in a hammerhead shape, which has not been observed among the top 10 SSI cases in the ERA5 dataset. This particular example indicates that some extreme samples generated by the standard GAN may seem unrealistic, and further studies are necessary to prove if such scenarios could happen. Despite these issues, the standard GAN achieves SSI values that are approximately 17.7% lower than those in the ERA5 dataset on average among the top 10 cases, indicating a relatively closer match in peak intensities.

Similarly, the WGAN-GP (Fig. 17) exhibits comparable spatial distribution, accurately identifying the key regions of high wind speeds (in brown and yellow). However, despite maintaining similar high-intensity patterns and areas, the SSI values of these top 10 cases are significantly lower, by approximately 46.8%. This suggests a consistent underestimation of extreme winds exceeding 24 m s$^{-1}$, leading to an overall underestimation in the SSI values.

The U-net diffusion model (Fig. 18) produces smoother and more consistent wind speed patterns across the maps while still capturing the general spatial characteristics of these extreme wind events. However, like the WGAN-GP, the generated maps of the top 10 cases show considerably lower SSI values, approximately 53.7% lower than the ERA5 dataset. This indicates a significant underestimation of peak intensities over 24 m s$^{-1}$.

Overall, while all models provide useful approximations of extreme wind scenarios, they exhibit limitations in accurately replicating the intensity and specific spatial details of these events. The standard GAN captures peak intensities closer to the ERA5 dataset but slightly falls short in spatial distributions. The WGAN-GP and U-net diffusion model, although showing better performance in generalising patterns, significantly underestimate peak intensities, impacting their effectiveness in representing extreme scenarios.

To further evaluate the performance of the models in capturing extreme wind scenarios, we analysed the top 100 SSI cases over land on the first two PC dimensions (Fig. 19). This analysis provides insights into how well the models replicate the distribution of extreme wind events in the reduced dimensional space.

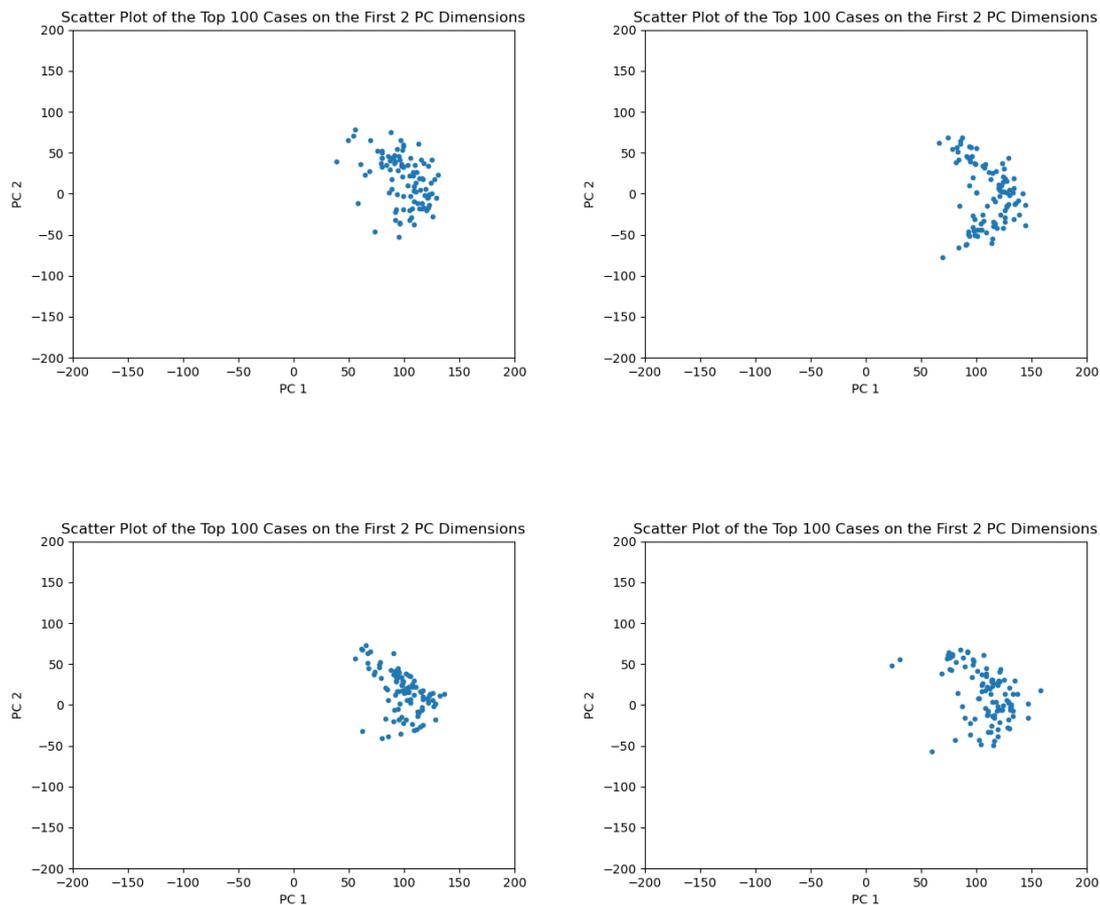

Figure 19. Scatter plots of the top 100 SSI cases on the first two PC dimensions for the ERA5 dataset (top left), standard GAN (top right), WGAN-GP (bottom left), and U-net diffusion model (bottom right).

The scatter plots for the top 100 SSI cases from the ERA5 dataset (top left) and the three models show similar clustering patterns, with PC1 ranging from 50 to 150 and PC2 ranging from -50 to 50. This similarity suggests that the models successfully capture the overall distribution of extreme scenarios. A notable observation is that the extremes in the data are along the PC1 dimension rather than the PC2 dimension. This indicates that the primary variability in extreme wind events is captured more significantly by the first PC, which can be associated with the overall intensity and spatial extent of windstorms. The second PC appears to capture variations that are less critical in defining extreme events, while further studies are necessary to identify if there is a different PC that also measures these extremes.

Subtle differences in the scatter plots highlight variations in model performance. The standard GAN (top right) shows a more dispersed cluster compared to the ERA5 dataset, especially on the PC2 dimension, where a few cases lie below -50. Moreover, while the extremes are located in roughly the same region among all four plots, the distribution from the standard GAN appears like a C-shape instead of a cluster-like shape. These observations indicate a different and slightly more variability in the extreme events that are not captured among the top 100 cases in the ERA5 dataset and the other two models.

On the other hand, the WGAN-GP (bottom left) displays a slightly tighter cluster with fewer outliers, suggesting a more consistent generation of extreme wind events. However, this tight clustering could also indicate a limitation in capturing the full variability of extreme scenarios, which could be due to the smaller sample size of the generated outputs compared to the actual ERA5 dataset.

The U-net diffusion model (bottom right) produces a cluster very similar to the ERA5 dataset, indicating a reasonable replication of the distribution of extreme events. However, the outliers from the U-net diffusion model lie slightly further from the main cluster, particularly on the PC1 dimension, with some points falling outside the range of [50,150]. This deviation might reflect the tendency of the model to generate smoother and more generalised wind speed patterns, which occasionally deviate from the observed extremes.

Overall, the scatter plots indicate that all three models are capable of capturing the general distribution of extreme wind events in the reduced dimensional space. The standard GAN shows more variability, the WGAN-GP produces a tighter but possibly less variable cluster, and the U-net diffusion model closely replicates the ERA5 dataset with some minor deviations.

*(iii)     Histograms of the Maximum SSI Values for Each Sample*

We also analysed the distribution of the logarithm with base 10 of the maximum SSI values across all grid points for each wind speed map (Fig. 20). The maximum SSI value for a map refers to the highest SSI value observed at any grid point within that specific sample. This value represents the peak intensities captured in every wind event, with the location of this peak value varying from sample to sample.

Given that many maps do not contain any wind speeds above the 98th percentile threshold, resulting in a maximum SSI value of zero, these zeros were removed to concentrate on non-zero extremes. To further improve visualisation, a logarithmic transformation to the non-zero maximum SSI values is applied to make it easier to compare the spreads between the models and the ERA5 dataset.

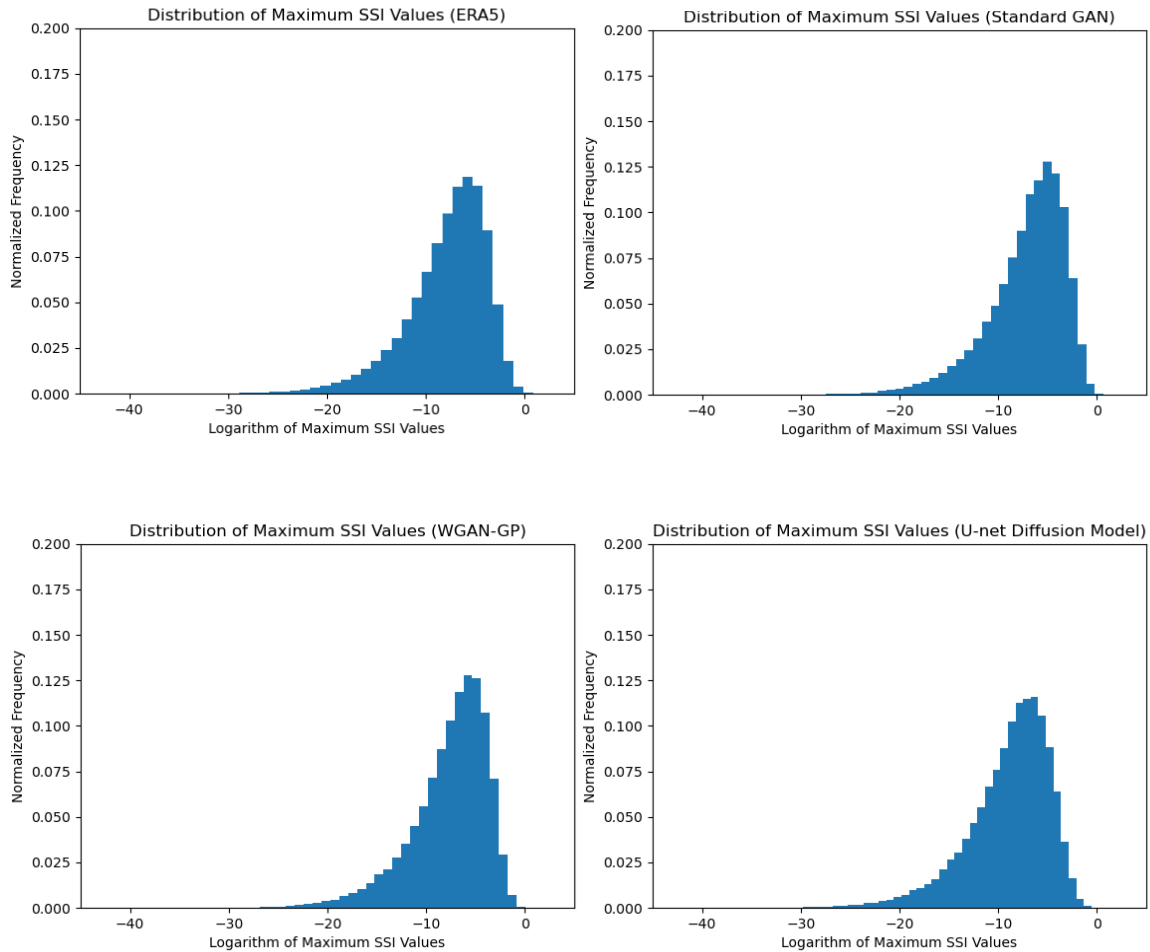

Figure 20. Normalised histograms of the logarithm with base 10 of maximum SSI values for the ERA5 dataset (top left), standard GAN (top right), WGAN-GP (bottom left), and U-net diffusion model (bottom right). Each histogram represents the distribution of the most intense storm risk captured at any grid point within each wind speed map.

The histogram of the ERA5 dataset reveals a negative skewness, where the tail extends towards the lower SSI values, indicating most wind speed maps generate moderately high logarithms of maximum SSI values. While all three models can capture the overall shape of the distributions, subtle differences are observed upon closer examination. The distributions for the WGAN-GP and U-net diffusion models show a slight shift towards more negative values on the x-axis by approximately one bin, implying these models may slightly underestimate the severity of extreme wind events. Moreover, the U-net diffusion model exhibits slightly less negative skewness (less steep on the larger end) compared to other distributions, indicating its tendency to generate extremes slightly less frequently.

*b.      Quantitative Evaluation of Model Performance*

| Metrics \ Dataset | ERA5 dataset | Standard GAN | WGAN-GP | U-net diffusion model |
|---|---|---|---|---|
| FID | N/A | 739 (2) | 721 (1) | 762 (3) |
| SSIM | 1.000 | 0.163 (1) | 0.155 (2) | 0.150 (3) |
| KL divergence | 0.000 | 1.41 (2) | 1.39 (1) | 1.45 (3) |
| EMD | 0.000 | 0.240 (2) | 0.219 (1) | 0.294 (3) |

Table 1. Performance summary of the generative models based on various metrics. Ranking are included in brackets, with (1) being the best performance for each metric. Metrics include the Fréchet inception distance (FID), structural similarity index measure (SSIM) of the average storm severity index (SSI) map, mean Kullback-Leibler (KL) divergence, and Earth mover's distance (EMD) on the first 25 principal component (PC) dimensions. The ERA5 dataset values are included as reference points (perfect scores).

The performance summary table (Table 1) compares the generative models across four different metrics. The WGAN-GP model consistently performs the best across most metrics, achieving the lowest FID, KL divergence, and EMD. This suggests that the WGAN-GP model produces images that are closest to the real dataset in terms of overall quality, variability, and distribution. However, it is slightly less accurate in replicating the spatial patterns in extreme scenarios, as reflected in its second place SSIM ranking, mainly due to the underestimations of wind speeds, as illustrated in Fig. 10.

The standard GAN performs well, particularly in SSIM. This indicates that the standard GAN generates images that most accurately replicate the spatial structures and intensities of the wind speed patterns in extreme scenarios. This may be due to the higher SSI values among these extremes compared to the other two models, which might maintain a closer resemblance to the ERA5 extremes. It also ranks second in other metrics, showing that it is a strong performer overall, though slightly behind the WGAN-GP in terms of overall quality and distribution.

The U-net diffusion model, while still effective, ranks third across all metrics. Its FID score suggests that it generates images with structures that are less similar to the ERA5 dataset compared to the other models. Additionally, its lower SSIM score indicates less spatial accuracy in replicating wind speed patterns in extreme scenarios, which aligns with the visual evaluation showing an underestimation of peak wind speeds. This underestimation leads to a lower magnitude in SSI, affecting the spatial patterns of the average SSI map. Its higher KL divergence and EMD score suggest that it is less effective in capturing the variability of the real data distribution.

Despite these differences, the models are indeed very close across all metrics, with no significant difference in the order of magnitude across a wide range of objective metrics. This indicates that all three generative models perform similarly well in generating wind speed maps, and the differences are relatively minor in the context of their overall performance. However, such metrics do not penalise models for generating potentially unphysical scenarios in a way that might be more readily detectable to domain experts. Therefore, in the next section, we survey meteorologists with a range of expertise to understand whether a more subjective metric may be useful.

*c.    Human Evaluation of Model Performance*

1)    SURVEY DESIGN AND METHODOLOGY

2)    SURVEY RESULTS

The results were analysed from three different perspectives: average realism scores for each model compared to the ERA5 dataset, average test scores for each proficiency level, and average realism scores for each range of SSI values. Fig. 21 shows the average realism scores for the ERA5 dataset and the three generative models, with higher scores representing greater realism. The ERA5 dataset received an average score of 2.64, which is slightly lower than the scores for all three models. The U-net diffusion model received the highest average score of 2.89, followed by the standard GAN at 2.84 and the WGAN-GP at 2.82. An average score of around 3 typically suggests respondents were unsure of the realism and chose a natural rating. This result shows that while the maps were able to confuse the respondents, the U-net diffusion model was perceived as generating the most realistic wind speed maps, slightly outperforming the other models.

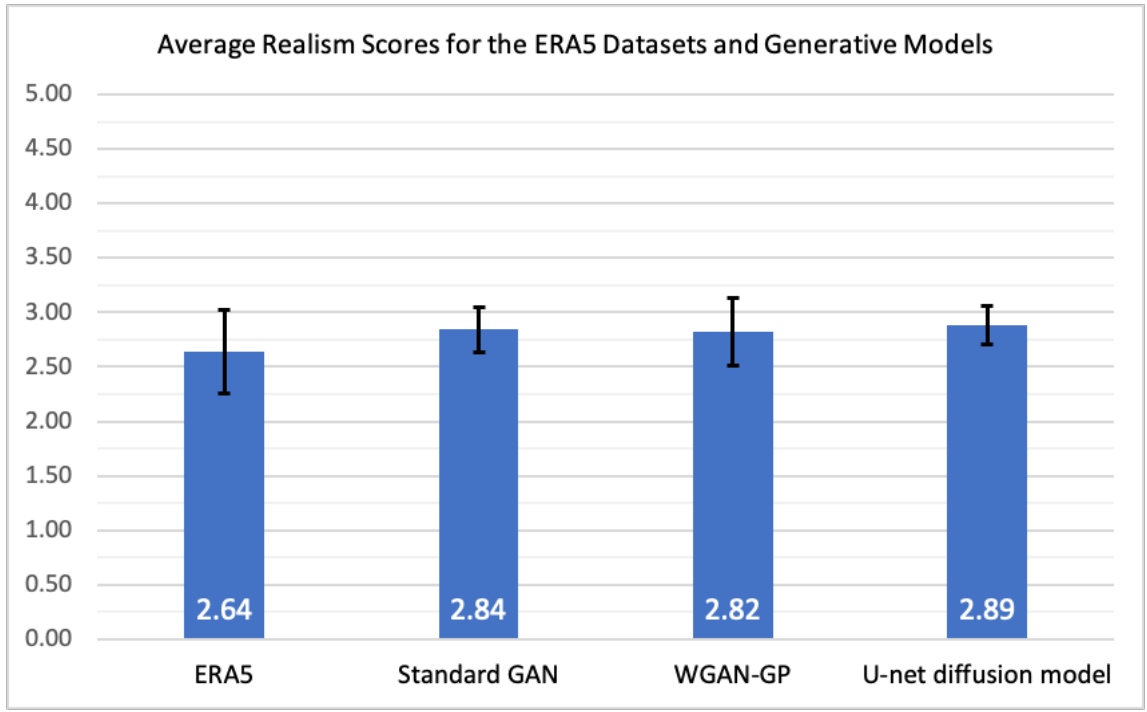

Figure 21. The average realism scores for the ERA5 dataset and the generative models (standard GAN, WGAN-GP, and U-net diffusion model). The scores indicate how realistic the wind speed maps from each model were perceived to be, with the highest scores representing greater realism. The black error bars represent the standard deviations of the scores among each map of the corresponding dataset.

Fig. 22 presents the average test scores based on the self-reported proficiency levels of the respondents. The scores were adjusted by defining ERA5 maps as 5 and the generated maps as 1, then calculating the average based on the difference from the perfect score. Interestingly, there is no sign that respondents with forecasting experience or a higher proficiency level in meteorology gave a higher average score. However, the highest average score of 3.35 was given by respondents with a PhD or research background in climate extremes or UK windstorms. This suggests that experts familiar with the data have a stronger ability to distinguish between actual reanalysis data and synthetic data generated by the models, yet they can still identify some differences. The higher score for this proficiency level may also be influenced by the smaller sample size of this group. Similarly, scores around 3 among all proficiency levels indicate that respondents chose a natural rating, which may be due to their uncertainties in distinguishing the maps.

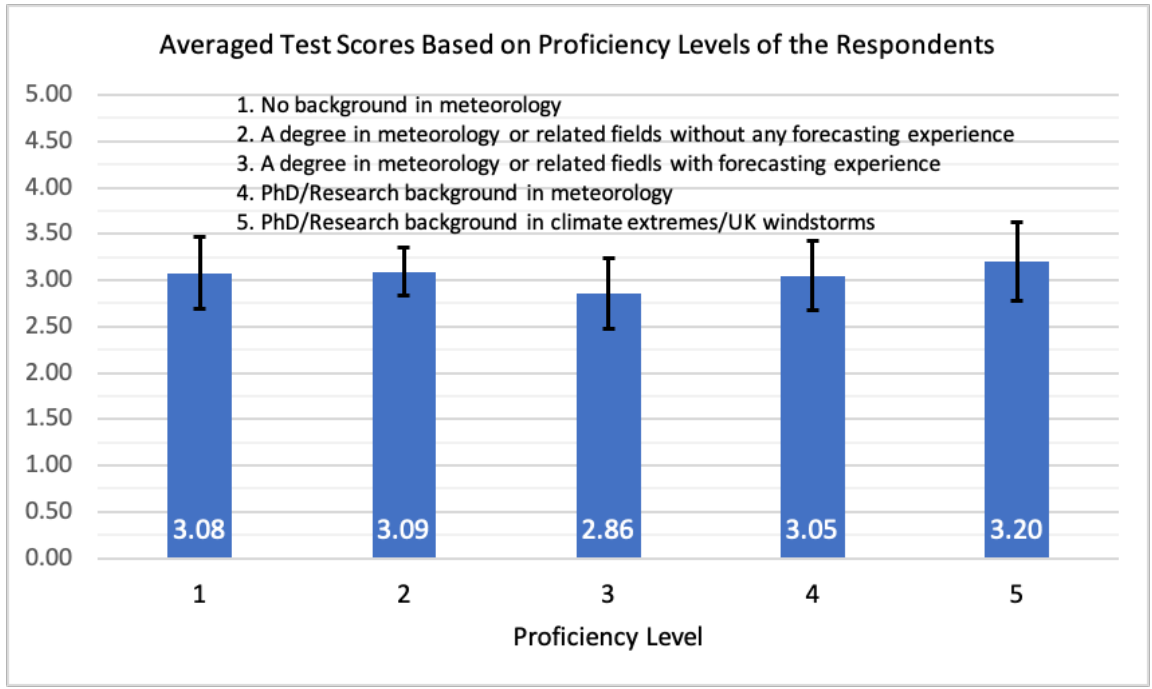

Figure 22. This plot presents the average test scores based on the self-reported proficiency levels of the respondents. The scores were adjusted by defining ERA5 maps as 5 and the generated maps as 1, then calculating the average based on the difference from the perfect score. The black error bars represent the standard deviations of the scores among each respondent of the corresponding category.

Fig. 23 illustrates the average realism scores for maps within different ranges of SSI values. Maps from the bottom 25% of SSI values received the lowest average score of 2.67, while those from the top 25% and the 25% to 75% range of SSI values received higher scores. These results suggest that respondents found maps representing moderate and severe wind speeds to be slightly more realistic than those representing the least severe scenarios. However, it could be due to a lack of distinct characteristics in calm wind scenarios, making it more difficult for respondents to distinguish the maps. As before, average scores close to 3 reflect the tendency of the respondents to choose a neutral rating.

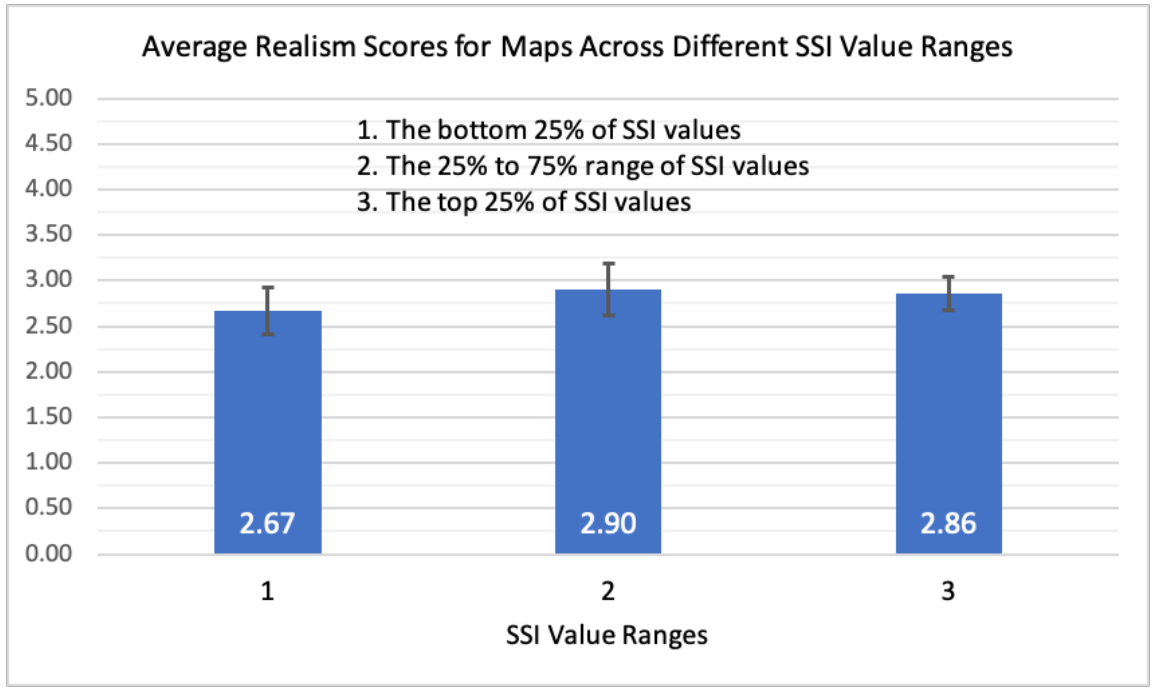

Figure 23. This plot illustrates the average realism scores for maps within different range of SSI ranges (the bottom 25% of SSI values, the 25% to 75% range of SSI values, and the top 25% of SSI values. The black error bars represent the standard deviations of the scores among each map of the corresponding category.

The survey results show that the U-net diffusion was generally rated as producing the most realistic wind speed maps, slightly outperforming the other two models. The fact that all average scores hover around 3 suggests that respondents often found it challenging to distinguish between real and synthetic maps, frequently opting for a neutral middle-ground rating. Respondents with specialised knowledge in climate extremes and UK windstorms provided the highest scores, while maps representing moderate and severe wind speeds were perceived as more realistic than those depicting the least severe scenarios.

However, it is important to recognise that a map looking realistic does not necessarily mean it is truly accurate. Factors like human biases, the rarity of certain patterns, and the clarity of spatial characteristics can all affect perceptions. The results can also be influenced by the small sample size. These findings serve as an additional assessment of the generative model capabilities and their effectiveness in producing realistic synthetic wind speed maps that will be crucial in guiding future improvements and refinements of these models.

# 5.    Discussions

*a.    Overview of Key Findings*

1)    SUMMARY

From the above evaluation, each model demonstrated a capacity to replicate the broad spatial characteristics of UK wind maps and successfully capture general patterns such as the separation between land and sea and the distinct wind behaviours across different regions. However, when these models were subjected to extreme windstorm scenarios, there were subtle differences in their performance.

The standard GAN model was noted for producing noisier images, which impacted the visual quality of its outputs. Despite this, the GAN model performed slightly better than the U-net diffusion model in the quantitative evaluations, suggesting that it is more effective in capturing the statistical properties of wind speed distributions. However, the slightly rotated PC1 vector from the standard GAN model compared to the ERA5 dataset and the other models suggests that there might have been subtle misalignments in how the standard GAN captured the primary modes of variation in the data. This rotation indicates that while the standard GAN managed to preserve variability, it might have introduced some distortions in representing the main patterns.

The WGAN-GP model showed improvements over the standard GAN in terms of the coherence and stability of the generated wind maps. It produced more coherent spatial structures and performed slightly better than the other models in the quantitative evaluations, suggesting a more accurate capture of the distributions and variability in wind patterns observed in the ERA5 dataset.

The U-net diffusion model stood out for its ability to generate wind speed maps with higher structural quality and less noise compared to the GAN-based models. The more stable and coherent representations likely contributed to its preference among human evaluators. However, the model showed a slightly weaker performance in capturing the statistical distributions and variability of wind speed, as assessed in the quantitative analysis. A slightly lower score in the SSIM also indicated the tendency of the model to underestimate extreme wind speeds, suggesting a trade-off where intense extremes are tuned down to achieve better generalisation and reduce noise.

A key observation across all evaluation methods was that the differences in performance among the models were often subtle, and the evaluation methods themselves did not always agree. For instance, while the U-net diffusion model was preferred in visual and human assessments mainly due to its noise-free outputs, quantitative metrics favoured the standard GAN and WGAN-GP models for their ability to capture statistical distributions more effectively. This disparity highlights that different aspects of model performance can be emphasised depending on the evaluation method used, and there is no single metric that fully captures all dimensions of model performance. The choice of evaluation method can significantly influence the model performance, indicating the complexity of evaluating generative models. These findings underscore the importance of combining various evaluation methods to obtain a comprehensive understanding of model performance.

In conclusion, while the U-net diffusion model produces visually coherent and noise-free images, it may slightly underperform in accurately capturing the full range of extreme wind speeds. On the other hand, the standard GAN and WGAN-GP models, despite producing noisier outputs, were more effective in capturing the statistical distribution of wind events, while the WGAN-GP performed slightly better in aligning with the ERA5 dataset. The differences highlight the importance of considering multiple aspects of model performance in terms of model selection and application.

## 2) MODEL PERFORMANCE ANALYSIS

Based on the summary of results, a deeper analysis reveals how the different methodologies influenced the performance of each generative model. This analysis will link the observed outcomes back to the specific characteristics of each model and the training process involved.

### (i) Standard GAN

In this study, efforts were made to stabilise the standard GAN through careful tuning of layer configurations and parameter settings. Multiple versions of the standard GAN architecture were trained, experimenting with different configurations of layers, learning rates, and batch sizes, to identify the optimal setup that minimises instability. The final version of the standard GAN likely avoided mode collapses due to a sufficiently deep architecture, the application of regularisation techniques, and the use of a large latent space input to represent variability (Shi et al., 2022). As a result, the evaluation highlighted the effectiveness of the model in capturing the overall statistical properties of wind speed distributions.

Despite these efforts, the model still exhibited some noise, which may have been due to the discriminator being occasionally outperformed by the generator, leading to less effective discrimination and nosier results. Additionally, the use of transposed convolutional layers in the generator that up-sample the tensors might have contributed to this observed noise. The slightly rotated PC1 vector in the generated outputs suggests a subtle misalignment in capturing the primary modes of variation, indicating that while variability was preserved, the representation of key patterns may have been slightly distorted.

*(ii)      WGAN-GP*

The WGAN-GP model is mainly built on the standard GAN architecture by incorporating a Wasserstein loss function and a gradient penalty, which were specifically chosen to address the stability issues observed in the standard GAN. The gradient penalty helped to stabilise the training process by ensuring that the discriminator provided more meaningful gradients, preventing the generator from diverging (Mescheder et al., 2018). Similar to the standard GAN, various configurations of the network layers and hyperparameters were tested to optimise performance, resulting in a model version that produced less noise and showed improved stability.

This enhanced stability is also reflected in its performance in quantitative evaluations, where the model slightly outperformed the other models. The model successfully captured the statistical distributions of wind speeds in both global and detailed perspectives, which is evidenced by its better alignment with the ERA5 dataset. Its ability to produce stable and quality outputs with fewer noises contributed to a better visual quality and accurate statistical representation.

*(iii)      U-net Diffusion Model*

The U-net diffusion model took a different approach, using a diffusion process to generate wind speed maps with smooth gradients and high visual quality. The U-net architecture was designed to capture both local and global features, and the diffusion process was implemented to ensure that noise was gradually added and removed, allowing the model to learn the data distribution effectively. Similarly, several configurations of the U-net layers and the diffusion process parameters were tested to select the best version.

This model performed the best in visual and human evaluations due to its aesthetically pleasing outputs. However, the model showed a slightly weaker performance in capturing the full range of variability, particularly in extreme wind scenarios. The diffusion process, while effective at reducing noise, may have contributed to an underestimation of extreme wind speeds, as indicated in the visual evaluation and the lower SSIM score of the average SSI map. This trade-off suggests that while the U-net diffusion model is excellent at generalising across typical scenarios, its focus on reducing noise might have smoothed out critical variations needed for accurately representing the most intense windstorm events.

## b. *Limitations of the Study*

### 1) DATASET LIMITATIONS

The spatial resolution constraints were introduced earlier to provide the necessary context for presenting the wind speed maps. Similarly, the sampling issues were discussed in the methodology section to introduce the inherent biases in the dataset due to the underrepresentation of certain extreme events. Temporal variability is also considered when interpreting the results. This section will focus on how other aspects of the dataset have influenced the research outcomes. These factors present additional challenges that impact the model ability to accurately simulate windstorm scenarios.

#### (i) *Temporal Inconsistencies*

The ERA5 dataset exhibits temporal inconsistencies that can impact the accuracy and reliability of the data. One significant source of these inconsistencies is the change in observational methods and technologies over the dataset coverage period from 1940 to 2022 (Hersbach et al., 2020). Prior to the satellite era, which began in 1979, observations were primarily collected from surface stations and sparse upper-air measurements. The introduction of satellite data significantly enhanced the accuracy and coverage of atmospheric observations, leading to a marked improvement in the quality of reanalysis data post-1979 (Bell et al., 2021).

These temporal discrepancies can lead to inconsistencies in the dataset, where pre-1979 data may exhibit higher uncertainty and potential biases compared to the more accurate post-1979 data. In the context of this study, generative models may recognise the inconsistencies between the two periods of data and potentially produce biased generated outputs.

The ERA5 dataset includes comprehensive historical meteorological data that reflects past climate conditions. This reliance on historical data introduces a significant bias, as the dataset is limited to patterns and events that have already occurred. Consequently, the models trained on this dataset are constrained by the characteristics of past windstorms and patterns, which may not fully represent the range of possible future scenarios, especially in the context of ongoing and future climate change.

One of the primary concerns is that historical data may underrepresent extreme events that could become more frequent or intense due to climate change (Konisky et al., 2016; Catto et al., 2019). Over the past decades, shifts in global climate systems have altered the distribution of atmospheric pressure systems, leading to changes in wind speed and direction (Trenberth, 1995). As climate continues to evolve, the wind patterns and behaviours of windstorms are expected to shift, potentially leading to scenarios that differ significantly from those captured in historical records (Peterson, 2000). The model reliance on historical data means they are less likely to generalise well to future conditions, leading to potential underestimations of risk for extreme weather events.

## 2)   METHODOLOGY LIMITATIONS

*(i)*     *Generative Model Suitability*

One of the main challenges with generative models or any ML models is their dependency on the quality and variety of the training data (Budach et al., 2022). In this study, the ERA5 dataset may lack sufficient examples of the most severe windstorm events, limiting the model ability to fully learn the characteristics of these rare occurrences. This limitation could lead to scenarios where the generated extremes are less intense than what might occur in reality, potentially underestimating the risks associated with future windstorms. The models might also generalise based on typical events, potentially smoothing out the most severe extremes in the generated outputs.

*(ii)*     *Limitations in Capturing Physical Dynamics*

Windstorms can be influenced by various interacting physical processes, including temperature gradients, pressure systems, and moisture content (Catto et al., 2019). Accurately simulating these dynamics is challenging but crucial for generating realistic windstorm scenarios, particularly for extreme events where small changes in physical conditions can lead to significant differences in outcomes (Lorenz, 1963). While the generative models are sophisticated in their ability to simulate spatial patterns in a purely statistical approach, they may not fully capture the intricate dynamics that drive windstorm formation and evolution. This limitation is due to the focus of the models on learning patterns rather than modelling the underlying physical processes. As a result, the generated windstorm scenarios may not always accurately reflect the true complexity of the atmosphere, particularly in extreme conditions where precise representation of physical interactions is critical.

*(iii)     Lack of Temporal Information*

Windstorms are not static events and their movement and intensification over time are critical factors that determine their overall impact. While the generative models used in this study create realistic snapshots of wind patterns, they cannot simulate the temporal evolution of these events. The absence of temporal information in the generated scenarios means that the models cannot simulate how a windstorm progresses, including where it moves, how it intensifies, and how it interacts with different geographical regions over time. However, the damage caused by windstorms is not only due to their intensity and spatial characteristics but also to their trajectory and duration (Della-Marta et al., 2009). Without the temporal dynamics, the generated scenarios may provide an incomplete picture of the potential impact of a windstorm.

*(iv)     Computational Resources Constraints*

The time required to train these models varies depending on their complexity. In the settings of this study, the standard GAN and WGAN-GP models each required approximately 30 minutes of training. However, the U-net diffusion model, which involves a more complex architecture and training process, required about 150 minutes. However, variations in model configurations and dataset size can significantly increase these training times, extending them to hours or even days, especially in the case of diffusion processes. These extended training periods can delay the research process and pose challenges for iterative experimentation to configure the model architectures and optimise the parameters. As the complexity of the models and the size of the datasets increase, the demands on computational resources grow exponentially. This constraint limits the ability to scale up the models for deeper layers, greater spatial coverage, or higher resolutions if possible.

The models also demand substantial RAM, both in terms of the central processing unit (CPU) and GPU. On the GPU side, RAM is particularly crucial during the training process, as it handles the large-scale operations and data manipulation required by deep learning models. Insufficient GPU RAM can slow down training exponentially or even cause the process to fail. For this study, although the models were trained with sufficient resources listed in the methodology section, the demands were still significant. On the other hand, CPU RAM is essential for generating samples after model training and for performing analyses on the dataset. High RAM requirements can be a limiting factor when generating high-resolution outputs and manipulating the datasets for further analysis.

*(v)*    *Black-box Nature*

Generative models operate as black boxes, meaning that these models generate outputs without providing clear, interpretable insights into how they arrived at those results. The internal processes within these models are highly complex and not easily understandable to humans (Fong & Vedaldi, 2017; Hassija et al., 2024). This black-box nature presents a limitation in understanding the underlying physical processes. Without transparency into how the model is making decisions, it becomes challenging to assess the reliability of the generated outputs and decrease user trust in the model.

Furthermore, the inability to interpret the inner workings of the models complicates the process of improving model performance. When a model produces unexpected results or fails to accurately simulate certain scenarios, it can be difficult to pinpoint the cause of the issue and make targeted adjustments (Papernot et al., 2017). This limits the effectiveness of the models in accurately capturing the statistical distributions or physical dynamics.

*(vi)*    *Challenges in Fine Tuning*

Fine-tuning generative models is a critical step in optimising their performance but this process presents significant challenges. One of the primary challenges is the intensive computational resources required as mentioned earlier. The need to repeatedly adjust parameters and retrain the models exacerbates these demands, making the process time-consuming and resource-intensive. The black-box nature of the model also makes fine-tuning a trial-and-error process, with no clear guidance on how to achieve the best possible results (Victoria & Maragatham, 2021).

Additionally, defining the correct optimisation objective and parameter space is another significant challenge, especially when the model must balance multiple factors. Implementing multi-objective optimisation schemes can address these challenges, but this approach exponentially increases computational demands and requires balancing competing goals (Deb, 2011; Cui et al., 2017). This consideration, alongside a large number of parameters within the model architecture and training process, adds another layer of complexity, which obscures how well the model is meeting each objective and the trade-off being made.

*(vii)    Choice of Evaluation Metrics*

As mentioned earlier, different evaluation metrics or methods often emphasise various aspects of the generated outputs. However, models that perform well on one metric may not necessarily perform well on others, leading to conflicting assessments of model performance. This discrepancy complicates the evaluation process, making it difficult to determine which models are truly performing better from a comprehensive perspective. Moreover, the specific requirements of simulating particular wind events introduce additional complexity. Traditional metrics such as FID and EMD may not adequately reflect the importance of extreme event simulation, which is critical for catastrophe modelling. As a result, introducing SSI-based metrics or performing additional evaluation on extreme cases are necessary for the intended application.

Human evaluation can also be a valuable tool for assessing model outputs, particularly in terms of visual and physical realism. However, this approach introduces subjectivity and requires a large number of respondents who are familiar with the ERA5 dataset for a comprehensive assessment. This introduces potential biases in the evaluation process, and the combination of subjective human assessments with objective quantitative metrics further complicates the assessment of model performance, as these evaluations may not always align.

*c.    Broader Implications*

1)    CONTRIBUTIONS TO THE FIELD

This research represents a significant advancement in the application of generative modelling in the field of meteorology. While generative models have been applied in areas such as nowcasting and climate downscaling, their use in generating synthetic meteorological datasets remains relatively unexplored, particularly in the UK. This study addresses this gap by demonstrating the potential of generative models, focusing on one of the most impactful meteorological events in the UK. By leveraging generative models, this study provides a novel approach to creating synthetic datasets that can be used to augment existing data, particularly rare extreme events, which are often underrepresented in historical datasets.

Moreover, the methodologies and findings from this research are transferable to other meteorological parameters and datasets. The techniques developed and validated in this study can be applied to simulate other weather-related events, which is crucial for enhancing data-driven predictive models that rely on large and diverse datasets to improve their accuracy and reliability. By providing a means to generate additional data points, particularly for rare events, this research contributes to the development of predictive modelling in meteorology.

2)    PRACTICAL APPLICATIONS

The generative models in this study have demonstrated considerable potential in simulating a wide range of wind events for the UK, which could create synthetic datasets to supplement existing data. However, the evaluation results indicate slight differences between the ERA5 dataset and the generated outputs, particularly the statistical distribution. Given the need for accurate and comprehensive datasets in CAT models, there remains room for improvement before they can be applied.

Nonetheless, this research offers a novel way to expand the available data, particularly for rare events that are underrepresented in historical records. Applying these datasets on a trial basis within CAT modelling could provide valuable insights into how the generative models perform under different conditions and identify areas for further refinement. Although the primary focus remains on CAT models in this study, this research lays the groundwork for broader applications as the technology and methodologies continue to advance.

# 6.    Conclusion

This research set out to explore the potential of generative models, specifically the standard GAN, WGAN-GP, and U-net diffusion model, in simulating wind speed maps for the UK, particularly extreme windstorms. Through a comprehensive evaluation of these models, the study has shown that while generative modelling is a promising approach for creating a wide range of realistic scenarios, several challenges and limitations remain.

One of the key findings is that the generative models developed in this study are capable of producing realistic spatial patterns and extreme windstorm scenarios. However, the evaluation also highlighted slight discrepancies between the ERA5 dataset and the synthetic datasets, particularly in terms of accurately representing the distribution of variability and the intensity of extreme events. These discrepancies suggested the importance of further refining the models to improve their accuracy and reliability, mainly when applied to risk assessments.

In conclusion, while the models have shown potential in generating realistic wind field maps, there is still room for improvement before they can be fully integrated into critical applications, particularly catastrophe modelling. Nevertheless, this study lays a strong foundation for future research and application, opening new avenues for the use of generative modelling to supplement existing dataset, specifically rare events, to improve the robustness of data-driven predictive models.

# 7.    Future Work

*a.    Exploring Model Architecture*

(1)    DEEPER NETWORKS

Future research could explore the development of deeper networks, including increasing the number of convolutional layers and trainable neurons and incorporating advanced architectural techniques. Increasing the depth and width of the network can allow the model to capture more complex patterns in the ERA5 dataset, particularly the tail-end patterns that represent extreme windstorms (Richter et al., 2021). Techniques such as self-attention and residual networks are also promising for enhancing the capability of deeper networks.

Self-attention mechanisms allow the model to dynamically focus on different parts of the input data, giving more weight to the most relevant information. It works by creating attention scores that measure the importance of each part of the input relative to others, enabling the model to capture long-range dependencies and complex interactions (Vaswani, 2017; Tang et al., 2018). In the context of this study, incorporating this technique can enhance the model ability to understand relationships between different regions of different wind events, such as how wind speeds in one area might influence conditions in another.

Residual networks (ResNets) are designed for very deep networks by introducing shortcut connections that skip one or more layers. These shortcuts allow the network to learn the residual, the difference between the output and the input of these skipped layers (He et al., 2016). This helps to prevent issues like vanishing gradients, where the gradients become too small as they are propagated through many layers (Borawar & Kaur, 2023). By maintaining strong gradient flows, ResNets allows the building of deeper networks that can learn more patterns without the degradation of performance. By incorporating these techniques into deeper networks, future research could enhance the model ability to simulate a wider range of wind events with greater precision.

(2)    NOVEL VARIATIONS

In addition to exploring deeper networks, future research could also consider novel variations in model architecture that combine the strengths of different approaches. One example is the diffusion GAN, which merges the strengths of GANs with the iterative diffusion process of diffusion models. This structure combines the use of an adversarial training framework to refine noisy inputs, potentially generating more stable and accurate outputs. This approach is exciting because it can reduce training time significantly while still maintaining high-quality results (Wang et al., 2022; Xiao et al., 2021). However, diffusion GANs are so novel that there are currently only two research papers available, making it an advanced area for further exploration.

Another example is the VAE-GANs, which combine the probabilistic latent space modelling of Variational Autoencoders (VAEs) with the adversarial training of GANs. The VAE component allows the model to learn a structured latent space, which captures the variations in the data (Xian et al., 2019). By playing with this latent space, future research can explore how different variations in the latent space influence the generated outputs. This approach enhances the diversity of the generated scenarios and provides a more interpretable framework for understanding how specific features and conditions affect the outputs.

*b.    Incorporating Temporal and Multidimensional Data*

(1)    MULTIDIMENSIONAL INPUTS

One way to expand the input data is to integrate various meteorological parameters into the models, such as temperature, pressure, and wind direction. By utilising a three-dimensional convolutional layer, the models can learn the complex correlation between these variables, leading to more realistic simulations in terms of physical processes (Tran et al., 2015). For instance, including information about temperature gradients, influence wind patterns, which could help models better capture the physical process driving storm formation and intensity in a purely data-driven approach. This multidimensional approach allows the model to reflect the interconnected nature of the atmosphere, improving their ability to simulate real-world conditions.

Another direction is to incorporate the time dimension by adding time as an additional dimension. The models can generate a time series of wind speed maps with three-dimensional convolutional layers, creating simulations that capture the evolution of weather events over time. Instead of simulating static snapshots of windstorms, this approach enables us to know their progression, intensification, and eventual dissipation. Such time series outputs would be helpful for risk assessments, where understanding the temporal dynamics of weather events is crucial.

(2)     FOOTPRINT APPROACH

While adding the time dimension directly as an additional input can create time series outputs, using footprints as the input data is a more practical and computationally efficient approach without many modifications to the models. It involves tracking the maximum wind speeds or wind gusts over specific periods, such as 24, 36, or 48 hours (Dawkins & Stephenson, 2018). This approach simplifies the process by providing a condensed summary of wind behaviours over time, capturing key temporal aspects, such as the path, speed, and intensity changes of windstorms. This method is useful in catastrophe modelling, where understanding the cumulative impact of a windstorm over time is critical for assessing potential damages (Dawkins, 2016).

*c.     Enhancing Application*

(1)     WIND GUST SIMULATION

As mentioned earlier, wind gusts have a more direct impact on structures than sustained wind speeds and are important in determining the potential damage during extreme windstorms. Further research should focus on refining the generative models to better simulate wind gust maps that directly contribute to catastrophe modelling.

(2)     NEW VARIABLES AND APPLICATIONS

Apart from wind speeds and wind gusts in this study, there is significant potential to extend the generative models to simulate other meteorological variables such as precipitation, snowfall, and temperature, which are crucial for various weather-related risk assessments. By expanding the range of variables, the models could focus on other meteorological phenomena, improving predictions related to floods, snow accumulations, heat waves, and other critical weather events. Additionally, applying these generative models to different geographical regions can test their generalisability in various climatic and geographical conditions. Extending the models to new datasets and regions will enhance their utility in a wide range of meteorological studies, supporting better risk assessments and decision-making in diverse fields.

## 1)    SAMPLING TECHNIQUES

As mentioned earlier, one key challenge is to ensure that the sampling for evaluation is consistent and representative. When comparing the generated outputs with the ERA5 dataset, discrepancies in sample sizes can lead to a biased assessment of model performance. The inclusion of temporal components in the ERA5 dataset introduces another layer of complexity in sampling. Future research could explore methods such as resampling to ensure that the sample sizes used for evaluation are equivalent while still representing a wide range of scenarios, allowing for a more accurate comparison between the datasets.

## 2)    POST-PROCESSING GENERATED OUTPUTS

The evaluation has suggested that the GAN-based models produce images with noises that may potentially influence the assessment of model performance. Fourier transform can be used to reduce high-frequency noise in the generated outputs and improve the realism and usability of the synthetic dataset (Wahab et al., 2021). By applying Fourier transform, the data is converted into the frequency domain, where high-frequency components that are often associated with noise can be isolated and filtered out (Barclay et al., 1997). This process preserves the dominant spatial patterns and essential structures, resulting in cleaner, more realistic outputs like the ERA5 dataset.

## 3)    EVALUATION METHODS

Evaluating the performance of generative models is challenging as it is crucial to balance different aspects of the generated outputs. While this study has employed a variety of metrics, there are additional complicated metrics that could further enhance the evaluation by focusing on different aspects of the generated data. One such metric is the normalised Laplacian pyramid distance (NLPD), which assesses the similarity by comparing deep feature representations across multiple scales and is useful for evaluating the fine-grained details in generated outputs (Ding et al., 2021). Another metric is the mutual information (MI), which measures the amount of information shared between two datasets (Kim et al., 2022). Adding evaluation metrics can achieve a more comprehensive understanding of the model performance.

Beyond quantitative metrics, human evaluation is also an important component of the assessment. Given the limited number of respondents in this study, future work could enhance this aspect by performing significance tests to validate the results and expanding the evaluation panel to include a more diverse group of experts. This will ensure that while the human evaluation is subjective, the results can still provide meaningful insights into the model performance.

*e.*     *Training across Time Periods*

Future research could explore training these models using datasets from different time periods to understand how climate change and temporal variations influence the generated outputs. For instance, comparing models trained on pre-industrial and mid-20[th] century could reveal how changing atmospheric conditions influence the intensity, frequency, and spatial patterns of windstorms. By assessing how well models trained on different time periods replicate recent or future conditions, we can investigate their potential or limitations in simulating future events under different climate change pathways.

Additionally, training across time periods can help identify the minimum dataset size and temporal range required to produce reliable models. Understanding these thresholds is crucial for optimising the training process, particularly when historical data is limited or when optimising the parameter configurations that require iterative training.

*f.*     *Hyperparameters Optimisation*

Hyperparameter optimisation is a critical aspect of improving the model performance, especially when dealing with complex datasets. Optuna is a power tool that can be leveraged for this purpose, which uses techniques such as Bayesian optimisation, grid search, and random search to efficiently explore the hyperparameter space (Akiba et al., 2019). By automating the search process and selecting promising hyperparameters such as learning rates, the number of layers, the size of latent spaces, and regularisation terms, it can enhance model performance with reduced computational resources. Future work should also consider the potential for multi-objective optimisation, where multiple performance criteria are optimised simultaneously. This requires extensive computational resources but ensures that the models perform well across a range of different aspects and weather scenarios of the generated outputs.

*Acknowledgements*

I would like to take the opportunity to express my deepest gratitude to everyone who has supported and guided me throughout this dissertation. Your meaningful insights and unwavering support have been invaluable.

First and foremost, I am grateful to my main supervisor, Dr. Kieran Hunt. I still remember our first conversation, where he enthusiastically presented over ten potential dissertation projects, each of which he had personally explored. His early provision of the dataset during the examination period, alongside his assistance in setting up the complex server and training environment, saved me a significant amount of time. Thank you for the consistent guidance during our weekly meetings, ensuring I remained on the right path, and for introducing me to this cutting-edge project that has been not only an incredible learning experience but also truly enjoyable.

I am also deeply thankful to Prof. Len Shaffrey for his insights into various aspects of climate research and analysis. His consistent encouragement was incredibly motivating and greatly reinforced my confidence throughout this journey.

My gratitude also extends to Prof. Atta Badii from the Department of Computer Science. Despite his limited background in meteorology, his engagement in our discussions and his insights into computational modelling offered a unique perspective that enriched my approach to the project. His participation in my dissertation presentation was particularly memorable and provided me with significant encouragement.

I would like to thank Dr. Richard Dixon from CatInsight and Dr. Ludovico Nicotina from Inigo Insurance for their expertise in catastrophe modelling and risk assessment. Their guidance provided me with the essential background and industry perspective necessary for this research.

To my family and friends, thank you for your support and encouragement throughout this journey. I am also grateful to my classmates, who fostered a warm and supportive environment. Working alongside each other, discussing our projects, and offering help when needed created a strong bond that I will always cherish.

Finally, while accurate weather prediction and risk assessment are vital, minimising the impact of catastrophes also requires a commitment to protecting our planet and mitigating climate change. As researchers, we have a responsibility not just to contribute to our fields but also to the well-being of the Earth.

*Data Availability Statement*

The dataset utilised in this study is the ERA5 hourly reanalysis data provided by the ECMWF, covering the periods from 1940 to 2022. It is publicly available and can be downloaded from the Copernicus Climate Data Store – "ERA5 hourly data on single levels from 1940 to present". This includes hourly wind speed and related meteorological variables necessary for the modelling and analysis in this study.

APPENDIX

In this appendix, we present the results of applying the same generative models used in the main analysis, including the standard GAN, the WGAN-GP, and the U-net diffusion model, to 10-m 3s wind gust instead of 10-m wind speed. The model architectures, parameters, and training processes remained identical, with only the input data being changed to observe how these models handle different meteorological variables.

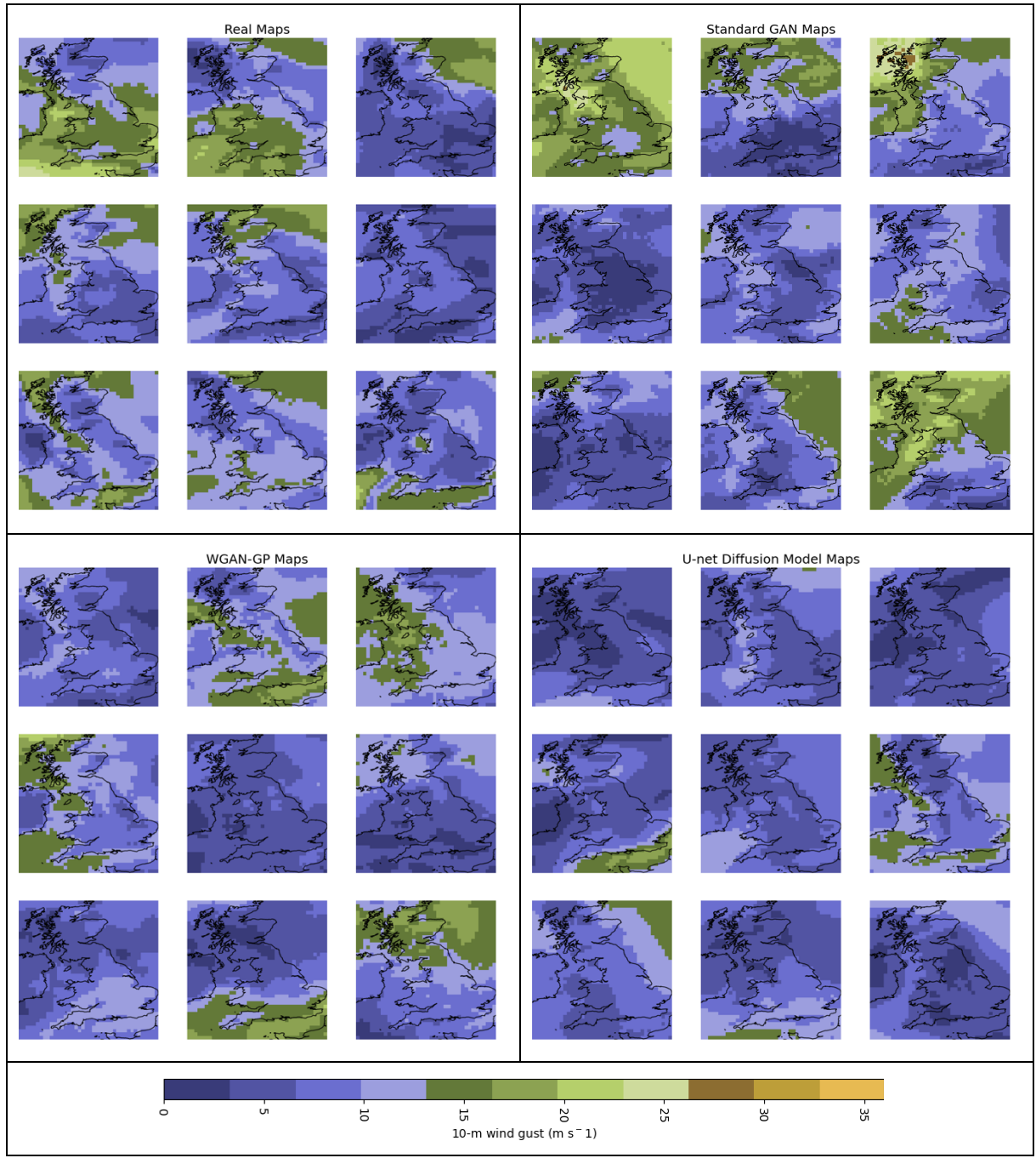

Figure 24. Comparison of typical wind gust maps. The plot shows nine random 10-metre 3-second wind gust maps in metres per second (m s⁻¹) from the ERA5 dataset (top left), standard GAN (top right), WGAN-GP (bottom left), and U-net diffusion model (bottom right).

Fig. 24 shows nine randomly selected maps from the ERA5 dataset and each model. The models maintained their ability in capturing the general spatial characteristics of the wind gust data observed in the ERA5 dataset. This finding suggests that these models are reasonably adaptable to varying input dataset in general without requiring alterations to their architecture, parameters, or training process.

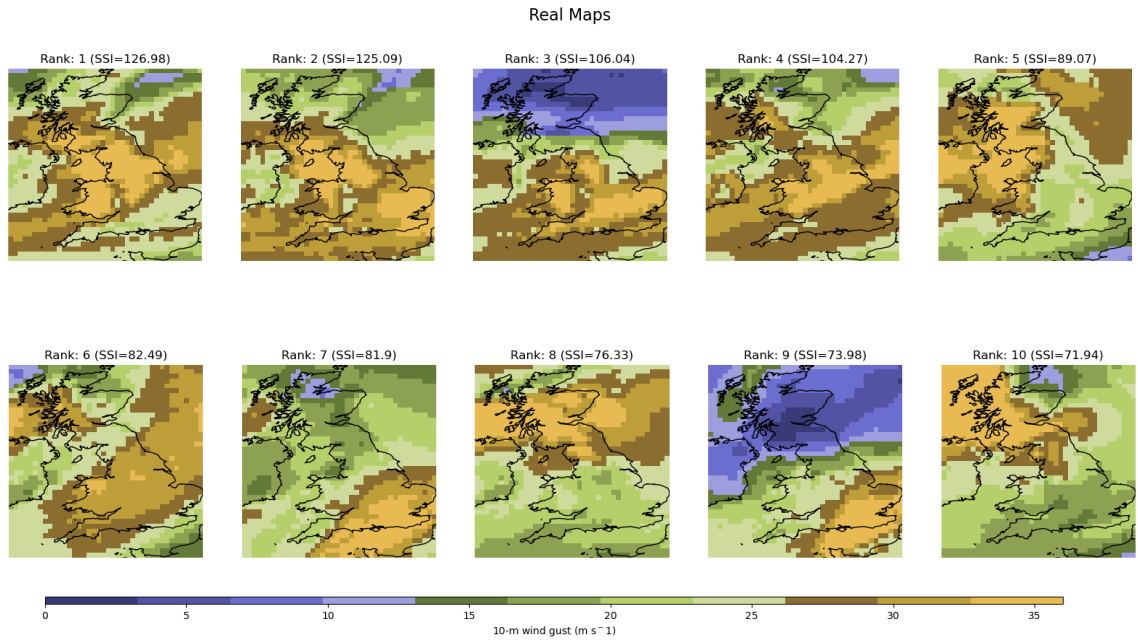

Figure 25. Top 10 SSI-ranked instances of extreme windstorm scenarios from the ERA5 dataset. The maps illustrate the 10-metre 3-second wind gust (in m s⁻¹) across the UK, showing areas with the highest wind gusts shaded in brown and yellow.

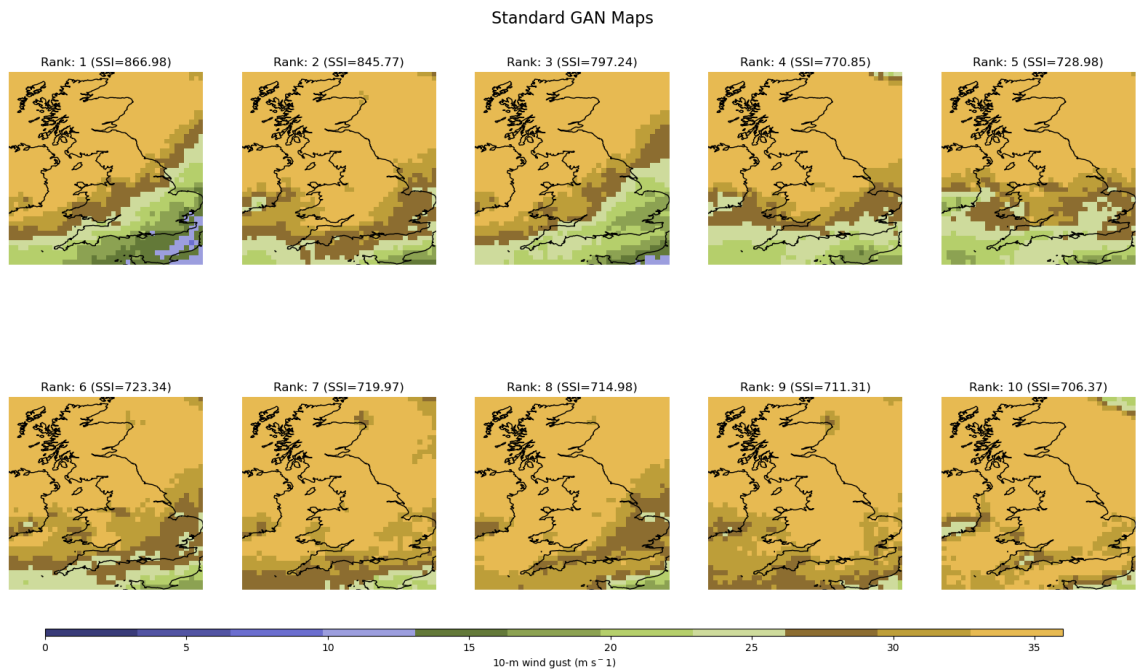

Figure 26. Top 10 SSI-ranked instances of extreme windstorm scenarios from the standard GAN. The maps illustrate the 10-metre wind speed (in m s⁻¹) across the UK, showing areas with the highest wind speeds shaded in brown and yellow.

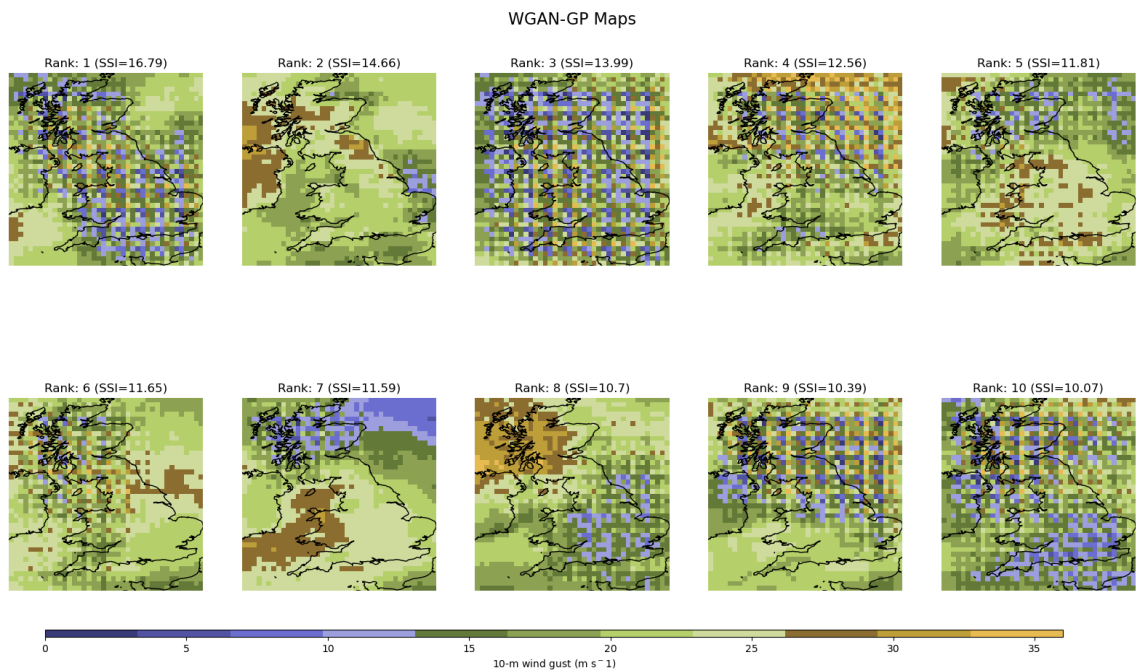

Figure 27. Top 10 SSI-ranked instances of extreme windstorm scenarios from the WGAN-GP. The maps illustrate the 10-metre wind speed (in m s⁻¹) across the UK, showing areas with the highest wind speeds shaded in brown and yellow.

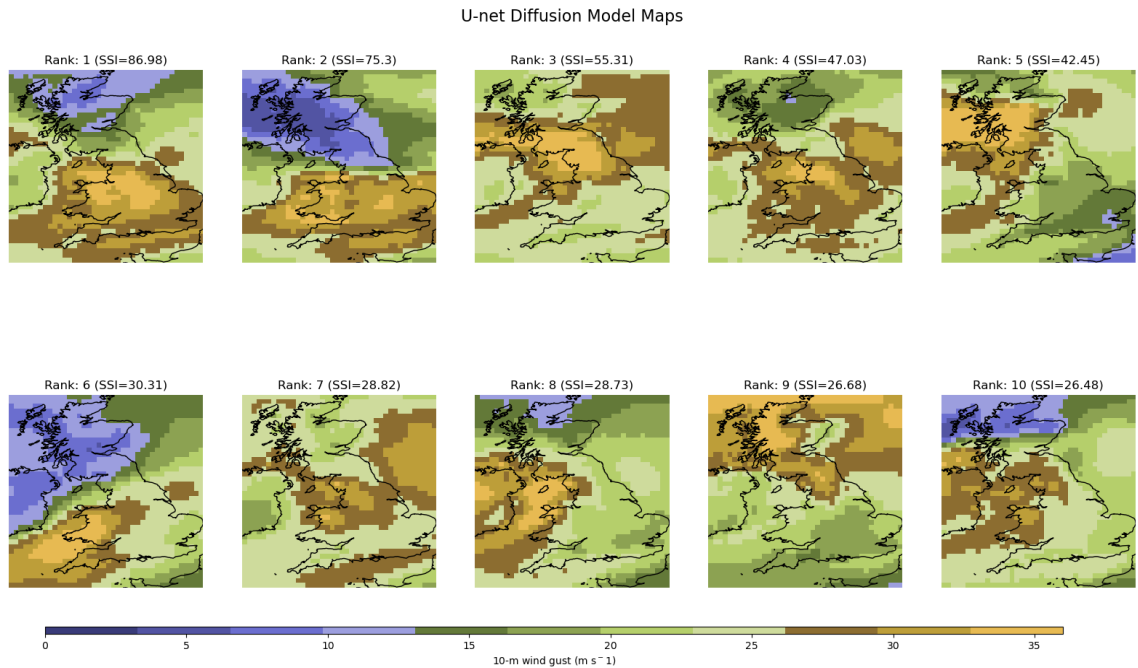

Figure 28. Top 10 SSI-ranked instances of extreme windstorm scenarios from the U-net diffusion model. The maps illustrate the 10-metre wind speed (in m s⁻¹) across the UK, showing areas with the highest wind speeds shaded in brown and yellow.

When evaluating the models under extreme scenarios, some models exhibited specific challenges in maintaining a consistent performance. The standard GAN (Fig. 26) faced difficulties in maintaining diversity in extreme scenarios, and significantly overestimated the intensities. This result indicates a potential sensitivity to the change in dataset.

On the other hand, the WGAN-GP model (Fig. 27) struggled with capturing the spatial characteristics, particularly the higher intensity regions of wind gusts. The output maps often showed blurred patterns, suggesting its inability to maintain the sharp gradients associated with extreme wind gust events.

The U-net diffusion model (Fig. 28) demonstrated a relatively consistent performance across both typical and extreme scenarios. The model successfully preserved the visual quality and spatial characteristics of extreme gusts. Its performance suggests that the U-net diffusion is less sensitive to changes in the input dataset compared to the other models, making it a more robust choice for handling different meteorological variables.

The results demonstrate the sensitivity of generative models to the characteristics of the input dataset, particularly the GAN-based models. The standard GAN and WGAN-GP models displayed vulnerabilities in accurately generating extreme events. This highlights the importance of careful tuning when applying generative models in different meteorological contexts.